\documentclass[aps,prd,preprint,superscriptaddress,nofootinbib]{revtex4-1}
\usepackage[utf8]{inputenc}
\usepackage{amsmath}
\usepackage{graphicx}
\usepackage{wasysym}

\makeatletter
\usepackage[usenames,dvipsnames]{xcolor}
\usepackage{slashed}
\definecolor{linkcolor}{RGB}{0,0,240}

\usepackage{hyperref} 
\hypersetup{
pdfstartview={XYZ},
colorlinks,
allcolors=linkcolor,
bookmarksopen
}

\makeatother

\begin{document}

\title{Relaxation leptogenesis, isocurvature perturbations, and the cosmic infrared background}

\author{Masahiro Kawasaki}

\affiliation{CRR, University of Tokyo, Kashiwa, 277-8582, Japan}

\affiliation{Kavli Institute for the Physics and Mathematics of the Universe (WPI),
University of Tokyo, Kashiwa, Chiba 277-8568, Japan}

\author{Alexander Kusenko}

\affiliation{Department of Physics and Astronomy, University of California, Los Angeles, CA 90095-1547, USA}

\affiliation{Kavli Institute for the Physics and Mathematics of the Universe (WPI),
University of Tokyo, Kashiwa, Chiba 277-8568, Japan}

\author{Lauren Pearce}

\affiliation{William I. Fine Theoretical Physics Institute, School of Physics and Astronomy, University of Minnesota, Minneapolis, MN 55455 USA}

\author{Louis Yang}

\affiliation{Department of Physics and Astronomy, University of California, Los Angeles, CA 90095-1547, USA}

\preprint{UMN-TH-3615/16,FTPI-MINN-16/35}

\begin{abstract}
Observations of cosmic infrared background (CIB) radiation exhibit significant fluctuations on small angular scales.  A number of explanations have been put forth, but there is currently no consensus on the origin of these large fluctuations.  We consider the possibility that small-scale fluctuations in matter-antimatter asymmetry could lead to variations in star formation rates which are responsible for the CIB fluctuations. We show that the recently proposed Higgs relaxation leptogenesis mechanism can produce such small-scale baryonic isocurvature perturbations which can explain the observed excess in the CIB fluctuations. 

\end{abstract}
\maketitle

\section{Introduction}
\label{sec:intro}

Observations of near-infrared cosmic infrared background (CIB) radiation by the AKARI and Spitzer space telescopes both have a consistent excess at the subdegree scale~\cite{Kashlinsky:2005di,2007ApJ...654L...5K,2012ApJ...753...63K,Cooray:2012xj,Seo:2015fga}. In particular, the integrated CIB fluctuation at 5 arcminutes, between 2 and 5 $\mu\text{m}$, is $\delta F_{2-5\,\mu\text{m}}\left(5^{\prime}\right)\simeq0.09\,\text{nW}\,\text{m}^{-2}\text{sr}^{-1}$~\cite{Helgason:2015ema,Kashlinsky:2014jja}.  This measurement of the anisotropic CIB entails that the power in the fluctuations is $F_\mathrm{CIB} \approx \delta F_\mathrm{CIB} \slash \Delta_{5^\prime} \sim 1 \; \mathrm{nW} \; \mathrm{m}^{-2} \; \mathrm{sr}^{-1} $.  The origin of this excess has not been clearly identified, but one plausible source is the first (population III) stars, which form at redshifts $z\gtrsim10$~\cite{Helgason:2015ema,Kashlinsky:2014jja}. 
While the AKARI observations can be explained by faint galaxies, the Spitzer observations are not consistent with this explanation~\cite{Helgason:2016xoc}.  (The Spitzer space telescope is able to resolve fainter point sources and does not observe a sufficiently large faint galaxy population to explain the excess~\cite{Helgason:2016xoc}.)  Zodiacal light is unable to account for the excess~\cite{Arendt:2016iph}.

The star formation rate depends on the distribution of halos, seeded by  cosmological density perturbations.  It was recently pointed out that, if primordial black holes account for dark matter, then isocurvature density perturbations arising from fluctuations in the distribution of black holes can explain the CIB measurements~\citep{ 2013ApJ...769...68C,Kashlinsky:2014jja,Helgason:2015ema,Kashlinsky:2016sdv}. In this scenario, the increase in the power of dark matter density perturbations on the small scales leads to a larger fraction of collapsed halos at redshift $z>10$.  This results in a higher $F_{\rm CIB}$, which can explain the CIB observations~\cite{Kashlinsky:2016sdv}. 
 
We here explore a different possibility.  Depending on its origin, the baryonic asymmetry of the universe can exhibit small-scale fluctuations.  These fluctuations can have the same effect on the CIB as the fluctuations produced by the black holes; namely, they can also increase the number of collapsed halos.   Models of ingomogeneous baryogenesis have been considered~\cite{Dolgov:1992pu, Dolgov:2008wu}.  In particular, the recently proposed Higgs relaxation leptogenesis models~\cite{Kusenko:2014lra, Pearce:2015nga, Yang:2015ida} are expected to produce small-scale baryonic isocurvature perturbations. A similar scenario can be constructed with other scalar fields, such as axions, or in models with an extended Higgs sector~\cite{Kusenko:2014uta,Gertov:2016uzs,Kusenko:2016vcq}.

This leptogenesis model is motivated by the observation that the Higgs field will generically undergo a post-inflationary relaxation epoch~\cite{Enqvist:2013kaa}.  Higgs relaxation leptogenesis uses an effective dimension $6$ operator in the scalar sector to produce an effective chemical potential during the Higgs relaxation epoch, which distinguishes matter from antimatter.  In the presence of a lepton-number-violating or baryon-number-violating interaction, the system relaxes towards its equilibrium state with nonzero asymmetry.

In the Higgs relaxation leptogenesis scenario, the final baryon asymmetry depends on the magnitude of the post-inflationary, pre-relaxation vacuum expectation value (VEV) of the Higgs field.  This can be produced by quantum fluctuations during inflation~\cite{Kusenko:2014lra,Yang:2015ida}.  Therefore this initial VEV, and consequently the produced asymmetry, will generically vary spatially.  In this work, we illustrate how these variations give rise to matter isocurvature perturbations.  Isocurvature perturbations are not affected by Silk or Landau damping, and baryonic isocurvature perturbations cannot be converted into adiabatic perturbations prior to the decoupling of baryons and photons \cite{Hu:1995xs}.  Therefore, such perturbations can cause massive regions to reach the non-linear regime earlier, enhancing star-formation at $z\simeq10$.  This provides an elegant resolution to the problem of excess CIB radiation.

This paper is organized as follows: In Section \ref{sec:Relaxation-Leptogenesis-Model}, we review the relevant features of the Higgs relaxation model and illustrate how it generates matter isocurvature perturbations.  Subsequently, in Section \ref{sec:Primordial-Baryonic-Isocurvature}, we calculate the spectrum of these baryonic isocurvature perturbations; we then consider how these modes evolve in Section \ref{sec:iso_evolution}.  The main results of this work are contained in Section \ref{sec:does_iso_work}, in which we show that these isocurvature modes cause sufficiently many halos large enough to support star formation to collapse around $z=10$ to explain the CIB observations.  Finally, we present the parameter space in which Higgs relaxation leptogenesis can both account for the observed matter-antimatter asymmetry of the universe and explain the CIB observations in Section \ref{sec:parameter_space}.

\section{The Higgs Relaxation Leptogenesis Model as a Source of Isocurvature Perturbations\label{sec:Relaxation-Leptogenesis-Model}}

In this section, we review the Higgs relaxation leptogenesis model, following the discussion in~\cite{Kusenko:2014lra,Yang:2015ida}, and then explain how it generates baryonic isocurvature perturbations.

During inflation, any scalar field $\phi$, including the Higgs field, with mass $m_{\phi}<H_{I}$ will develop a vacuum expectation value (VEV) $\sqrt{\left\langle \phi^{2}\right\rangle }$ through quantum
fluctuations~\cite{Linde:1982uu,Starobinsky:1982ee,Vilenkin:1982wt}.  Due to Hubble friction, the field is unable to efficiently relax to its equilibrium value.  The average VEV can be computed via a stochastic approach, which we discuss in detail below.  At the end of inflation, the Hubble parameter decreases, and the scalar field will relax to its equilibrium value.

For successful Higgs relaxation leptogenesis, we additionally assume that the Higgs field is coupled to the $(B+L)$ fermion current, $j_{B+L}^{\mu}$,
through an operator of the form 
\begin{equation}
\mathcal{O}_{6}=-\frac{1}{\Lambda_{n}^{2}}\left(\partial_{\mu}\left|\phi\right|^{2}\right)j_{B+L}^{\mu},
\label{eq:O6operator}
\end{equation}
which can be arranged by coupling $\phi^2$ to $-g^{2}W_{\mu\nu}^{a}\tilde{W}_{a}^{\mu\nu}+g^{\prime2}B_{\mu\nu}\tilde{B}^{\mu\nu}$ and using
the electroweak anomaly equation, among other possibilities~\cite{Kusenko:2014lra,Pearce:2015nga,Yang:2015ida}. As the VEV of $\phi$ evolves in time, this operator acts as an effective chemical potential, 
\begin{equation}
\mu_{\mathrm{eff}}=\frac{1}{\Lambda_{n}^{2}}\partial_{t}\left|\phi\right|^{2},
\end{equation}
for the fermion current $j_{B+L}^{\mu}$.  In the presence of a $B$ or $L$-violating interaction (such as those mediated by heavy right-handed neutrinos), the system will acquire a nonzero $B+L$ charge.  The available parameter space was described in Ref.~\cite{Yang:2015ida}; here we simply emphasize that this included regions of parameter space in which the right-handed neutrino is too heavy to thermalize, thus suppressing thermal leptogenesis.  The final lepton-number-to-entropy ratio is in general determined by the initial VEV $\phi_{0}$ at the end of inflation, $Y \propto \phi_0^2$, as explained in Appendix \ref{ap:analytical}.

We emphasize that since the effective chemical potential $\propto \partial_t |\phi|^2$, it is independent of the phase of $\left< \phi \right>$, and therefore, the same sign asymmetry is generated in all Hubble patches.  Consequently, it is not necessary for the observable universe to be contained within one Hubble patch.  Due to quantum fluctuations, these different regions of the universe will generically have different initial VEVs $\phi_{0}$ right after the inflation.  Since the asymmetry is proportional to the initial VEV, different patches in the universe will end up with different baryon asymmetries after the above-described leptogenesis mechanism is completed.  As time progresses, different scales will re-enter the horizon; as baryons become non-relativistic, these baryonic density fluctuations will evolve, and some may collapse.  The observable universe today consists of many Hubble patches, and therefore we expect enhanced baryonic fluctuations in the Higgs relaxation leptogenesis model.

Since the Higgs field $\phi$ is not the inflaton, and we ensure that it does not dominate the energy density of the universe, the baryonic fluctuations generated in this manner are isocurvature (entropy) perturbations.  They are independent from the adiabatic (curvature) perturbations produced during reheating by the decay of the inflaton.  

This production of baryonic isocurvature perturbations in Higgs relaxation leptogenesis was noted in Ref.~\cite{Kusenko:2014lra,Yang:2015ida}, where it was observed that these perturbations have the potential to exceed observational bounds from the cosmic microwave background radiation (CMB)~\cite{Planck:2013jfk}.  Therefore, these isocurvature perturbations must be suppressed at scales probed by the CMB.  This led to the construction of the ``IC-2" initial condition in those references, in which the Higgs field $\phi$ is massive ($m_{\phi}>H_{I}$) at the beginning of the inflation, due to a coupling to inflaton via one or several operators of the form
\begin{equation}
\mathcal{L}_{\phi I}=c\frac{\left(\phi^{\dagger}\phi\right)^{m/2}\left(I^{\dagger}I\right)^{n/2}}{M_{pl}^{m+n-4}}.
\label{eq:H_Inflaton_Coupling}
\end{equation}
While the inflaton VEV $\left\langle I\right\rangle $ is large, these operators provide a large effective mass to the scalar field $\phi$, suppressing the growth of its VEV due to quantum fluctuation. As inflation proceeds and $\left\langle I\right\rangle $ decreases, the Higgs field $\phi$ becomes effectively massless ($m_{\phi}<H_{I}$), and the vacuum expectation value starts to grow. As we discuss below, the initial VEV, and therefore the resulting asymmetry, depends on $N_{\mathrm{last}}$, the number of $e$-folds (measured from the end of inflation) that the Higgs VEV developed during.  In references~\cite{Kusenko:2014lra, Yang:2015ida,Gertov:2016uzs}, we set $N_{\mathrm{last}} \sim 8$ out of an abundance of caution; next, we discuss more precisely the exact observational constraint.

\section{Spectrum of Primordial Baryonic Isocurvature Perturbations\label{sec:Primordial-Baryonic-Isocurvature}}

Having explained how the Higgs relaxation leptogenesis model produces baryonic isocurvature perturbations, we now proceed in this section to determine the spectrum of these primordial baryonic isocurvature perturbations.  We will also apply observational constraints to the spectrum, and we will determine how this constrains $N_\mathrm{last}$, the number of $e$-folds the Higgs VEV grows during.  

We will first need to calculate the spectrum of the fluctuations in the Higgs vacuum expectation value, since this sources the fluctuations in the baryon density.  As mentioned above, in Higgs relaxation leptogenesis models, the Higgs field is coupled to the inflaton in such a way that the vacuum expectation value grows during only the last $N_{\mathrm{last}}$ $e$-folds of inflation.  If the effective mass turns off sufficiently fast, then the average VEV of $\phi$ at the end of inflation in a completely flat potential is 
\begin{equation}
\phi_{0}\equiv\sqrt{\left\langle \phi^{2}\right\rangle }\approx\sqrt{\int_{H_{I}e^{-N_{\mathrm{last}}}}^{H_{I}}\frac{dk}{k}\left(\frac{H_{I}}{2\pi}\right)^{2}}=\sqrt{N_{\mathrm{last}}}\frac{H_{I}}{2\pi}.
\end{equation}
This ignores curvature in the potential; a more accurate determination of the VEV is found by first solving the Fokker-Planck equation \cite{Starobinsky:1994bd}
\begin{equation}
\frac{\partial P\left(\phi,t\right)}{\partial t}=\frac{\partial^{2}}{\partial\phi^{2}}\left[\frac{H_{I}^{3}P\left(\phi,t\right)}{8\pi^{2}}\right]+\frac{\partial}{\partial\phi}\left[\frac{P\left(\phi,t\right)}{3H_{I}}\frac{dV\left(\phi\right)}{d\phi}\right],
\end{equation}
for $P\left(\phi,t\right)$, the probability distribution function
of observing the VEV equal to $\phi$ at time $t$.  [$V\left(\phi\right)$ is the potential for the scalar $\phi$; in this case, our scalar is the Higgs boson.]  The time evolution of the average VEV of $\phi$ can then be computed through 
\begin{equation}
\left\langle \phi^{2}\left(t\right)\right\rangle =\int d\phi\;\phi^{2}P\left(\phi,t\right),
\label{eq:real_phi0}
\end{equation}
with the initial condition $P\left(\phi,t=0\right) = \delta\left(\phi\right)$.  In our analysis, we make use of the Higgs potential at one loop, with running couplings where the RG equations are calculated at two loops, following~\cite{Degrassi:2012ry}.  We use the same potential, with thermal corrections, to evaluate the post-inflationary relaxation of this vacuum expectation value, as in~\cite{Kusenko:2014lra,Pearce:2015nga,Yang:2015ida}.  $\phi_0$ denotes the vacuum expectation value at the end of inflation, which is the initial VEV for the Higgs relaxation epoch.

This vacuum expectation value is produced by quantum fluctuations, and therefore it is not constant in space, as was mentioned above.  Perturbations are produced on all physical spatial scales inside the horizon  $l\lesssim H_{I}^{-1}$, where the Hubble parameter is evaluated when the VEV begins to grow (that is, $N_{\mathrm{last}}$ $e$-folds before the end of inflation).  Therefore, perturbations exists in all of the subhorizon modes which have physical momentum $p=k/a>H_{I}$.  As the modes exit the horizon ($p=k/a\lesssim H_{I}$), these perturbations become classical and are frozen with the amplitude
\begin{equation}
\delta\phi_{k}\equiv\Delta_{\phi}\approx\frac{H_{I}}{2\pi}
\end{equation}
per unit interval in $\ln p/H_{I}$ \cite{Linde:2005ht}. The isocurvature perturbations are approximately conserved in the superhorizon regime because the Higgs field does not contribute significantly to the energy density.

We define $k_{s}=a(N_{\mathrm{last}})p_{s}\sim a(N_{\mathrm{last}})H_{I}$, the comoving wavenumber corresponding to the mode which leaves the horizon as the fluctuations in the Higgs field are first produced.  The power spectrum of $\phi$ is then approximately
\begin{equation}
\mathcal{P}_{\phi}\left(k\right)\approx\begin{cases}
0 & \text{for }k < k_{s},\\
\left(\frac{H_{I}}{2\pi}\right)^{2} & \text{for }k \ge k_{s}.
\end{cases}\label{eq:Pc_phi(k)}
\end{equation}
In principle, one can further determine the details of the power spectrum from the transition from the Higgs field from an effectively massive field to an effectively massless field, which depends on the specific form of the operators \eqref{eq:H_Inflaton_Coupling} which couple the Higgs to the inflaton, generating the large effective mass during the early stages of inflation.  

As discussed in Sec.~\ref{sec:Relaxation-Leptogenesis-Model}, these perturbations in the Higgs VEV $\phi$ generate isocurvature perturbations in the baryon asymmetry $Y_{B}$.  These perturbations have a spectrum 
\begin{equation}
\left.\frac{\delta Y_{B}}{Y_{B}}\right|_{k} = \frac{\delta\left(\phi^{2}\right)_{k}}{\left\langle \phi^{2}\right\rangle } \approx \frac{2\ln^{1/2}\left(k/k_{s}\right)}{N_{\mathrm{last}}}\theta\left(k-k_{s}\right),
\label{eq:YB_ini}
\end{equation}
up to a large scale cutoff; see Appendix \ref{ap:power_spectrum_Y}. This makes use of the improved analytical estimates in Ref.~\cite{Gertov:2016uzs}; see the discussion in Appendix \ref{ap:analytical}.  We note here that the CIB signal will be dominated by $k \approx 1.4 k_s$, as we will discuss in Section \ref{sec:does_iso_work}.  As the universe cools, this induces a baryon energy density perturbation with the same spectrum
\begin{equation}
\delta_{B}\left(k\right)\equiv\left.\frac{\delta\rho_{B}}{\rho_{B}}\right|_{k}=\left.\frac{\delta Y_{B}}{Y_{B}}\right|_{k}.\label{eq:dB_ini}
\end{equation}

Having determined the spectrum, we now consider observational constraints.  For scales $k\apprle0.1\,\text{Mpc}^{-1}$, measurements of the cosmic microwave background radiation (CMB) from the Planck and WMAP collaborations constrain the baryonic isocurvature perturbation~\cite{Planck:2013jfk}.  The measured upper bound on the completely uncorrelated isocurvature fraction is given by
\begin{equation}
\beta_\mathrm{iso} = \dfrac{\mathcal{P}_{\mathcal{SS}}(k_{*p}) }
{\mathcal{P}_{\mathcal{SS}}(k_{*p})+\mathcal{P}_{\mathcal{RR}}(k_{*p})}
\end{equation}
where $\mathcal{P}_{\mathcal{RR}}$ is the power spectrum of the adiabatic fluctuation, $\mathcal{P}_{\mathcal{SS}}$ is the power spectrum of the isocurvature fluctuation, and $k_{*p}$ is the pivot scale used by the Planck collaboration.  Planck reports bounds evaluated at three momentum scales for a variety of models (see Table 15 of Ref.~\cite{Ade:2015lrj}).  To constrain our model, we use the most conservative bound from the CDI general model, making use of TT, TE, EE, low P, and WP data:
\begin{align}
\beta_\mathrm{iso}(k_{*p} = 0.002 \, \mathrm{Mpc}^{-1}) \lesssim 0.021, \nonumber \\
\beta_\mathrm{iso}(k_{*p} =0.050 \, \mathrm{Mpc}^{-1}) \lesssim 0.034, \nonumber \\
\beta_\mathrm{iso}(k_{*p} =0.100 \, \mathrm{Mpc}^{-1}) \lesssim 0.031. 
\end{align}
Since we are interested specifically in the baryonic isocurvature perturbation, we rescale the power spectrum by a factor of $(\Omega_b \slash \Omega_{DM})^2$.  Thus the requisite bound is:
\begin{align}
\left|\frac{\delta Y_{B}}{Y_{B}}\right| \lesssim \dfrac{\Omega_{DM}}{\Omega_b} (\beta_\mathrm{iso} \mathcal{P}_{\mathcal{RR}})^{1 \slash 2},
\end{align}
where $\mathcal{P}_{\mathcal{RR}}^{1 \slash 2} \approx 2.2 \times 10^{-9}$~\cite{Ade:2015lrj}.  This gives constraints of $|\delta Y_B \slash Y_B| \lesssim 3.4 \times 10^{-5}$ at $k_{*p} = 0.002 \, \mathrm{Mpc}^{-1}$, $4.3 \times 10^{-5}$ at $k_{*p} = 0.050 \, \mathrm{Mpc}^{-1}$, and $4.1 \times 10^{-5}$ at $k_{*p} = 0.100 \; \mathrm{Mpc}^{-1}$.  However, these constraints may be evaded by taking $k_s > 0.100 \; \mathrm{Mpc}^{-1}$, which corresponds to producing isocurvature perturbations on scales smaller than those probed by Planck.  Observations of the primordial spectrum in the CMB data at these scales are limited by the Silk (photon diffusion) damping.

At smaller scales, $0.2\,\text{Mpc}^{-1}\lesssim k\lesssim10\,\text{Mpc}^{-1}$, the Lyman-$\alpha$ forest provides information on the matter power spectrum, which strongly restricts isocurvature perturbations~\cite{Beltran:2005gr}.  Again, we will evade this bound by taking $k_s \gtrsim 10 \, \mathrm{Mpc}^{-1}$.  We note that despite the large comoving momentum, these isocurvature perturbations remain cosmologically relevant as isocurvature perturbations are not affected by Silk damping \cite{Hu:1995xs}.

Next, we connect $k_s$ to $N_\mathrm{last}$, the number of $e$-folds during which the Higgs VEV grows.  The results given below are exact in the limit that the curvature of the potential is negligible.  In our parameter space plots in section \ref{sec:parameter_space}, we use similar reasoning with the exact calculation of the initial Higgs VEV in a curved potential, using equation \eqref{eq:real_phi0}.

The mode that is exiting the horizon $N_{\mathrm{last}}$
$e$-folds before the end of inflation (that is, the mode that corresponds to $k_s$) grows to a size of $l_{\mathrm{EOI}} \simeq e^{N_{\mathrm{last}}}H_I^{-1}$ at the end of inflation (EOI).  Subsequently during reheating, the scale factor  $a$ grows by a factor of
\begin{equation}
\frac{a_{RH}}{a_{\mathrm{EOI}}}=\left(\frac{\Lambda_{I}}{T_{RH}}\right)^{4/3},
\end{equation}
where $\Lambda_{I}$ is the energy scale of inflation and $T_{RH}\approx\left(3/\pi^{3}\right)^{1/4}g_{*}^{-1/4}(T_{RH})\sqrt{m_{pl}\Gamma_{I}}$
is the reheat temperature. After reheating, the entropy of the universe
is conserved, 
\begin{equation}S=a^{3}s=2\pi^{2}g_{*s}(T)a^{3}T^{3}/45,
\end{equation} 
which allows us to relate the current scale factor to the scale factor at the end of reheating,
\begin{equation}
\frac{a_{\mathrm{now}}}{a_{RH}}=\frac{g_{*S}^{1/3}(T_{RH})}{g_{*S}^{1/3}(T_{\mathrm{now}})}\frac{T_{RH}}{T_{\mathrm{now}}},
\end{equation}
where $T_{\mathrm{now}}=2.73\,\text{K}$, the effective number of
relativistic species is $g_{*S}(T_{RH})=106.75$ for $T>300\mathrm{\,GeV}$,
and $g_{*S}(T_{\mathrm{now}}) = 43/11$ for $T=T_{\mathrm{now}}$ (in the Standard Model).  Combining these relations, the mode that exits the horizon $N_{\mathrm{last}}$ $e$-folds before the end of inflation corresponds to a perturbation mode with the comoving momentum 
\begin{equation}
k \simeq 2\pi e^{-N_{\mathrm{last}}}H_{I}\left(\frac{T_{RH}}{\Lambda_{I}}\right)^{4/3}\frac{g_{*S}^{1/3}(T_{\mathrm{now}})}{g_{*S}^{1/3}(T_{RH})}\frac{T_{\mathrm{now}}}{T_{RH}},
\label{eq:k_st_and_N_last}
\end{equation}
where we have set the scale $a_{\mathrm{now}}=1$, so that the comoving wavenumber coincides with the physical wavenumber now; thus $k = 2\pi \slash \mathcal{\ell}_\mathrm{now}$.

\begin{figure}
\includegraphics[width=0.48\columnwidth]{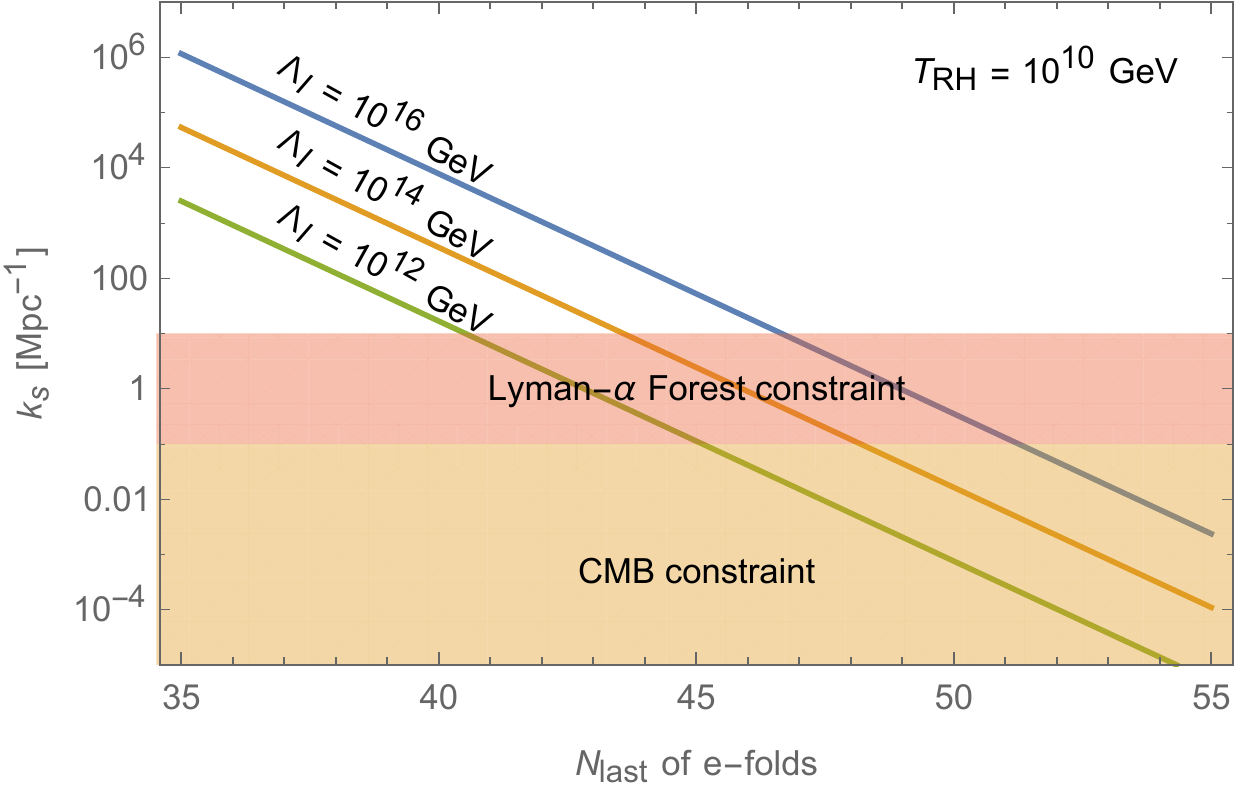}\includegraphics[width=0.48\columnwidth]{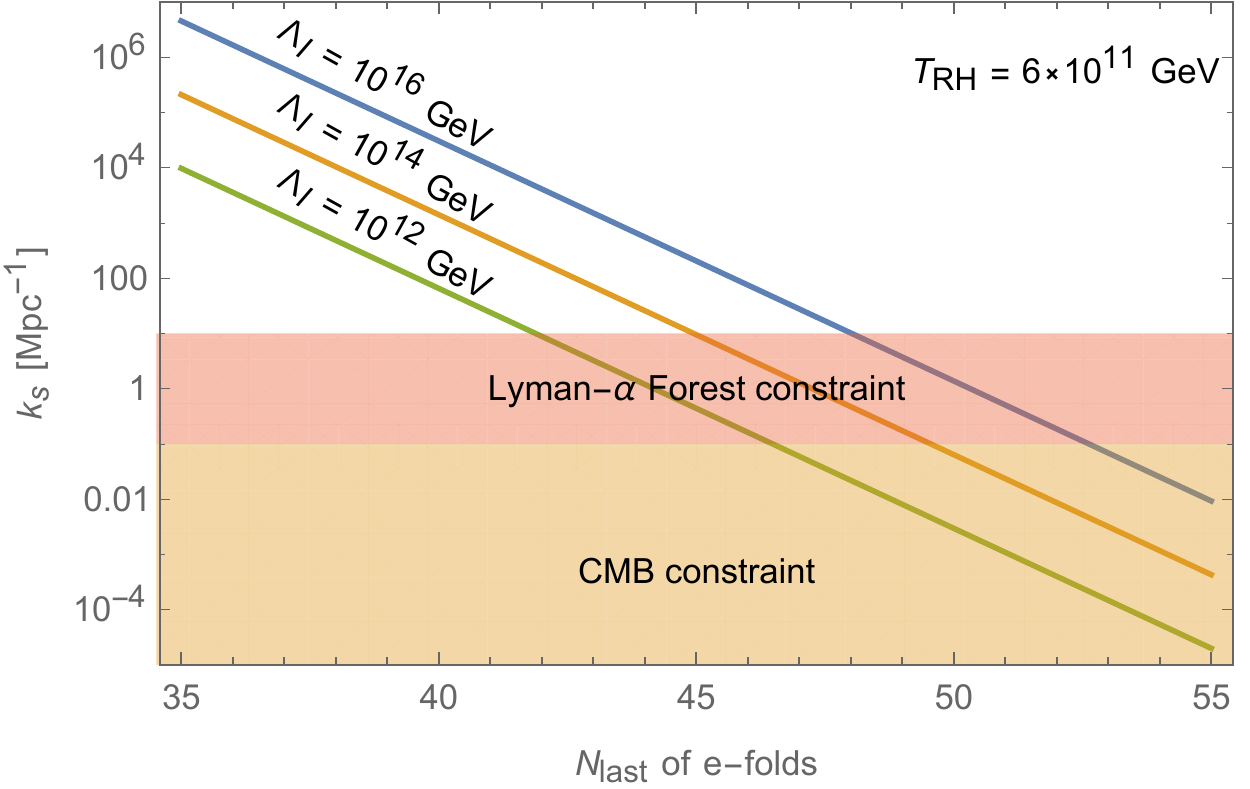}

\caption{The solid lines show $k_s$ as a function of $N_\mathrm{last}$, using equation \eqref{eq:k_st_and_N_last}, for various values of the inflationary scale $\Lambda_I$ and reheat temperature $T_{RH}$.  
The orange (red) region indicates the constraints on $k_s$ from the CMB (Lyman-$\alpha$ forest) observations.
\label{fig:Nlast vs ks}}
\end{figure}

Therefore, the requirement that isocurvature perturbations are generated at scales $k_s \apprge k_{*}=10\,\text{Mpc}^{-1}$,
which corresponds to a limit on $N_{\mathrm{last}}$ of
\begin{align}
N_{\mathrm{last}} &\apprle
48.2 -\ln\left(\frac{k_{*}}{10\,\text{Mpc}^{-1}}\right)+\frac{2}{3}\ln\left(\frac{\Lambda_{I}}{10^{16}\,\text{GeV}}\right)+\frac{1}{3}\ln\left(\frac{T_{RH}}{10^{12}\,\text{GeV}}\right)\nonumber \\
& \qquad +\frac{1}{3}\ln\left(\frac{g_{*S,\,\mathrm{now}}}{3.91}\right)
-\dfrac{1}{3} \ln \left( \dfrac{g_{*S,RH}}{106.75} \right)
+\ln\left(\frac{T_{\mathrm{now}}}{2.73\,\text{K}}\right),
\label{eq:N_last_limit}
\end{align}
which is not very stringent. The allowed parameter space for baryonic isocurvature perturbations is illustrated in Fig.~\ref{fig:Nlast vs ks}. The restrictions on $k_s$ from the CMB and Lyman-$\alpha$ forest discussed above can be converted into limits on $N_\mathrm{last}$ through the use of \eqref{eq:k_st_and_N_last}; these are also shown in Fig.~\ref{fig:Nlast vs ks}.  

We note that the Lyman-$\alpha$ forest constraints apply to the total contribution from both adiabatic and isocurvature perturbations.  We recall that adiabatic perturbations have $\mathcal{R}=\sqrt{A_{S}}\cong4.7\times10^{-5}$ if one assumes a flat spectrum.  $N_{\mathrm{last}} \approx 40 \sim 50$ corresponds to an initial baryonic density contrast of $\delta_{B,\,0} \approx 0.02 \sim 0.03$ at $k = 1.4 k_s$, using equations \eqref{eq:YB_ini} and \eqref{eq:dB_ini}.  This entails that the baryonic isocurvature perturbations generally dominate the adiabatic perturbations in the range where both are present.  Therefore, as Fig.~\ref{fig:Nlast vs ks} shows, it is indeed necessary to impose that $k_s \gtrsim 10 \, \mathrm{Mpc}^{-1}$.

\begin{figure}
\includegraphics[scale=0.6]{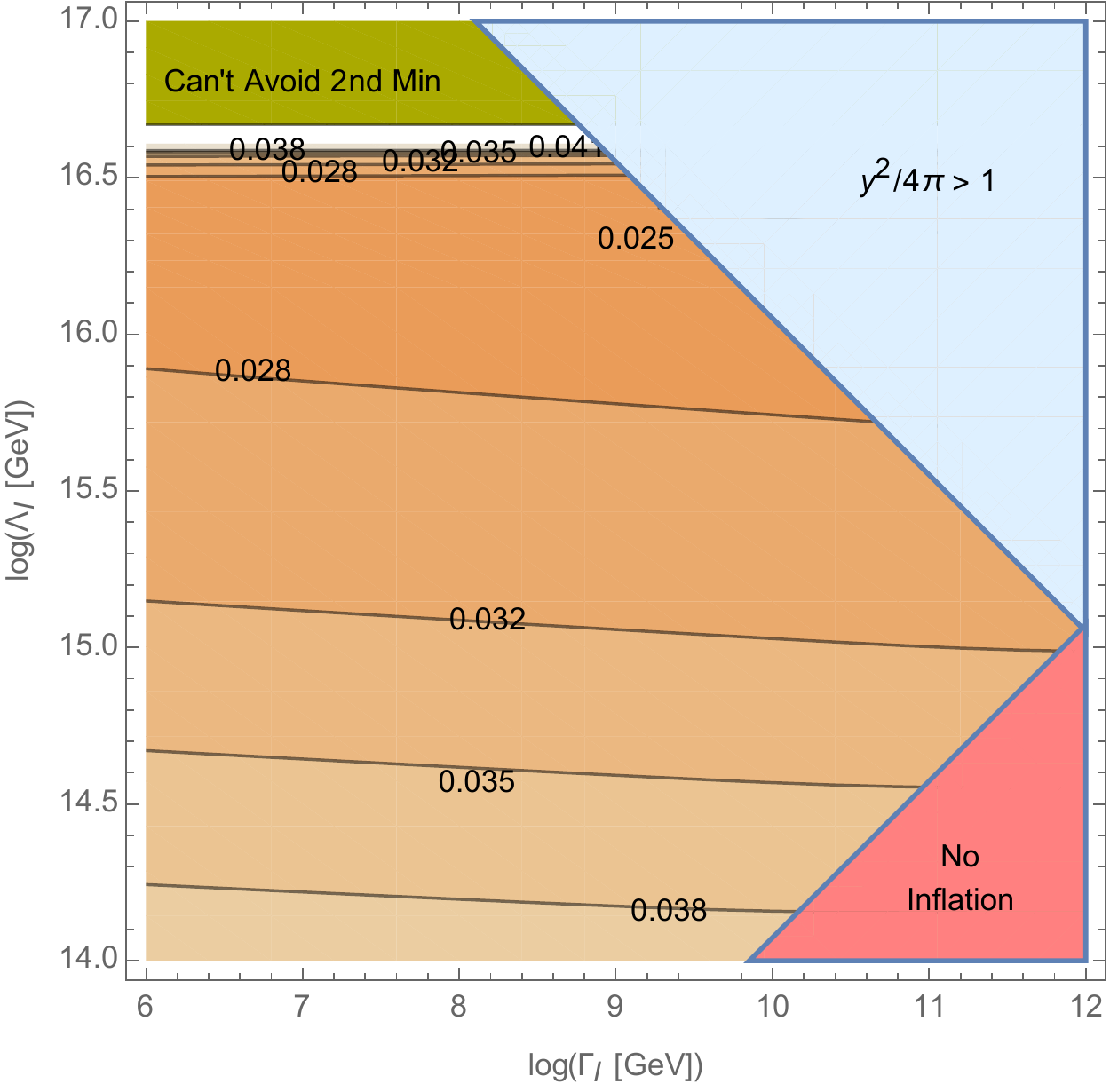}
\caption{The variation of $\delta_{B,\,0}(k= 1.4 k_s)$ in parameter space, with $k_s = 65 \, \mathrm{Mpc}^{-1}$.  In the green region on the upper left, the $N_\mathrm{last}$ given by equation \eqref{eq:k_st_and_N_last} is large enough that the Higgs VEV probes the minimum of the Higgs potential at large VEVs.  (Details on the parameters used in the calculation the potential can be found in~\cite{Yang:2015ida}.) 
In the red region on the lower right, $\Gamma_I > 3 H_I$ so  inflation doesn't happen.  As in Ref.~\cite{Yang:2015ida,Gertov:2016uzs} we set the neutrino Yukawa coupling such that right handed neutrino mass, inferred from the seesaw mechanism, is large enough that thermal leptogenesis is insufficient to explain the observed baryon asymmetry; in the upper right hand corner (light blue region), this would lead to a non-perturbative coupling.  We see that there is a slight variation in $\delta_{B,\,0}$ over the available parameter space.}
\label{fig:delta_B_parameter_space}
\end{figure}

In fact, we will see in section \ref{sec:does_iso_work} that we best explain the CIB with $k_s \approx 65 \, \mathrm{Mpc}^{-1}$.  We note here that this corresponds to the perturbations beginning to grow around 46.5 $e$-folds before the end of inflation with $\Lambda_I = 10^{16}\,\text{GeV}$ and $T_{RH} = 10^{12}\,\text{GeV}$.  From equations \eqref{eq:YB_ini} and \eqref{eq:dB_ini}, this corresponds to an initial baryonic density contrast of $\delta_{B,\,0} \approx 0.025$ at $k = 1.4 k_s$.  However, the second equality in equation \eqref{eq:YB_ini}, which was used with \eqref{eq:dB_ini}, holds in the limit of a flat potential.  Accounting for the curvature in the potential, using \eqref{eq:real_phi0}, decreases $\phi_0$, and so consequently increases $\delta Y_B \slash Y_B$ slightly.  In Fig.~\ref{fig:delta_B_parameter_space}, we have fixed $k_s = 65 \, \mathrm{Mpc}^{-1}$ and used equation \eqref{eq:k_st_and_N_last} to solve for the appropriate $N_\mathrm{last}$ and $\phi_0$ at each point in parameter space.  We then calculated $\delta_{B,\,0}$ at $k = 1.4 k_s$ at each point.  (We recall that, as mentioned above, this will be the scale most relevant to explaining the CIB excess.)  As expected, $\delta_{B,\,0}$ is slightly enhanced as compared to the flat potential case; this becomes more pronounced as $\Lambda_I$ decreases.\footnote{As noted, in the green region at the top left, the Higgs VEV probes the global minimum at large VEV values (see \cite{Enqvist:2013kaa}).  As this region is approached, the Higgs VEV explores the ``hilltop'' that divides the two minima, where the potential becomes flat.  Therefore, increasing $\Lambda_I$ leads to a larger increase in $\phi_0$, and consequently, the denominator of $\delta_B$ grows at a faster rate.  It grows faster than the numerator, which scales as $H_I$.  This accounts for the decrease in $\delta_B$ in the top left of the figure.\vfill}

To summarize the results of this section, the Higgs relaxation model generates baryonic isocurvature perturbations with a spectrum given by Eq.~\eqref{eq:dB_ini}.  The single free parameter in the spectrum, $k_s$, can equivalently (for fixed $\Lambda_I$ and $\Gamma_I$) be taken to be $N_\mathrm{last}$, the number of $e$-folds before the end of inflation during which the Higgs VEV grows.  (However, since $N_\mathrm{last}$ affects the VEV $\phi_0$, this then influences the final asymmetry produced by Higgs relaxation leptogenesis.)  By taking $k_s > 10 \, \mathrm{Mpc}^{-1}$, or (approximately) equivalently, $N_\mathrm{last} \lesssim 48$, the isocurvature perturbations evade all current observational bounds.

\section{Evolution of the Baryonic Isocurvature Perturbations}
\label{sec:iso_evolution}

In the previous sections, we explained how the Higgs relaxation model produces isocurvature perturbations, and we found the spectrum of these isocurvature perturbations.  Next, we consider the evolution of the isocurvature perturbations during the subsequent evolution of the universe.  We note that due to the tight coupling between photons and baryons, the amplitude of the isocurvature baryonic perturbations $\delta_{B}$ does not evolve before photon decoupling at $z\approx1100$.  (In fact, this was implicitly used above when we imposed constraints from the observations of the cosmic microwave background radiation and the Lyman-$\alpha$ forest.)

To study the late-time spectrum of the baryonic isocurvature perturbation, we calculate the evolution of the perturbations using the linearized Einstein equations and the linearized equation from  conservation of the energy-momentum tensor.  We work in the conformal Newtonian gauge, in which the scalar metric perturbation is parameterized as
\begin{equation}
ds^{2}=a^{2}\left(\tau\right)\left[\left(1+2\Phi\right)d\tau^{2}-\left(1-2\Phi\right)d\boldsymbol{x}^{2}\right].
\end{equation}
In our analysis, we consider the following components: radiation
(denoted by $i=r$), dark matter ($i=DM$), and baryons ($i=B$). 
The equations of state are parameterized by $w_{r}=1/3$ for radiation and $w_{DM}=w_{B}=0$ for baryons and dark matter (that is, we consider cold dark matter).  We assume that dark matter does not support sound waves, $u_{s,\,DM}^{2}=0$, and we make the tight coupling limit that baryons and photons share the same velocity potential $v_{B}=v_{r}\equiv v_{Br}$ before decoupling.  Therefore the effective speed of sound squared for the baryon and radiation fluids is $u_{s,\,Br}^{2}=1/3\left(1+R_{B}\right),$ where $R_{B}=3\rho_{B}/4\rho_{r}$.  However, we do not impose $4\delta_{B}=3\delta_{r}$, which is appropriate only for adiabatic modes.  

Therefore, the complete system of equations describing the evolution of the perturbations prior to recombination is \cite{Gorbunov:2011zzc}
\begin{align}
k^{2}\Phi+3\mathcal{H}\Phi'+3\mathcal{H}^{2}\Phi & =-\frac{a^{2}}{2M_{pl}^{2}}\sum_{i}\rho_{i}\delta_{i},\\
\delta'_{DM}-k^{2}v_{DM} & =3\Phi',\\
\delta'_{B}-k^{2}v_{Br} & =3\Phi',\\
\delta'_{r}-\frac{4}{3}k^{2}v_{Br} & =4\Phi',\\
v'_{DM}+\mathcal{H}v_{DM} & =-\Phi,\\
v'_{Br}+\mathcal{H}\frac{R_{B}}{1+R_{B}}v_{Br}+\frac{3}{4}u_{s,\,Br}^{2}\delta_{r} & =-\Phi,
\label{eq:linear_theory}
\end{align}
where $\mathcal{H}\equiv a^{\prime}/a$ and a prime denotes the derivative with respect to the conformal time defined via $d\tau=dt/a\left(t\right)$. 
The Hubble parameter in cosmic time, $t$, and in conformal time, $\tau$, are related by $H(t) = \mathcal{H}(\tau)/a $, and the Hubble parameter can be well described by $H = H_0 \sqrt{\Omega_{m}/a^3 + \Omega_{r}/a^4 + \Omega_{\Lambda}}$ with $a = 1 \slash (1+z)$ after the universe enters radiation domination. The density perturbation spectra $\delta_i$ generically have both isocurvature and adiabatic contributions.

After recombination at $z\approx1100$, photons and baryons decouple
and so $v_{B}$ and $v_{r}$ evolve separately. The perturbation equations
for baryons and radiation are then replaced by 
\begin{align}
\delta'_{B}-k^{2}v_{B} & =3\Phi',\\
\delta'_{r}-\frac{4}{3}k^{2}v_{r} & =4\Phi',\\
v'_{B}+\mathcal{H}v_{B} & =-\Phi,\\
v'_{r}+\frac{1}{4}\delta_{r} & =-\Phi.
\end{align}

For large scales $k<k_{s}$, we assume the initial density perturbation spectra $\delta_i$ satisfy the adiabatic conditions
\begin{equation}
\delta_{DM,\,0}=\delta_{B,\,0}=\frac{3}{4}\delta_{r,\,0}=-\frac{3}{2}\Phi_{0}=\mathcal{R},
\end{equation}
with a scale invariant spectrum. The Planck 2015 data set gives $A_{S}=e^{3.089}10^{-10}$ at $k=0.05\,\text{Mpc}^{-1}$ \cite{Ade:2015lrj}, which corresponds to the initial amplitude $\mathcal{R}=\sqrt{A_{S}}\cong4.7\times10^{-5}$.

For small scales $k>k_{s}$, we include the baryonic isocurvature
perturbations in addition to the adiabatic perturbations. For the parameters of interest, the isocurvature contribution to $\delta_{B,\,0}$ will generally dominate over the adiabatic contribution, and therefore $\delta_{B,\,0}\left(k\right)$ is given by Eq.~\eqref{eq:dB_ini}.  For the other components, we take $\delta_{DM,\,0}=\frac{3}{4}\delta_{r,\,0}=-\frac{3}{2}\Phi_{0}=\mathcal{R}$
for $k > k_S$, since these have only the adiabatic contribution.

\begin{figure}
	\includegraphics[width = 0.6 \columnwidth]{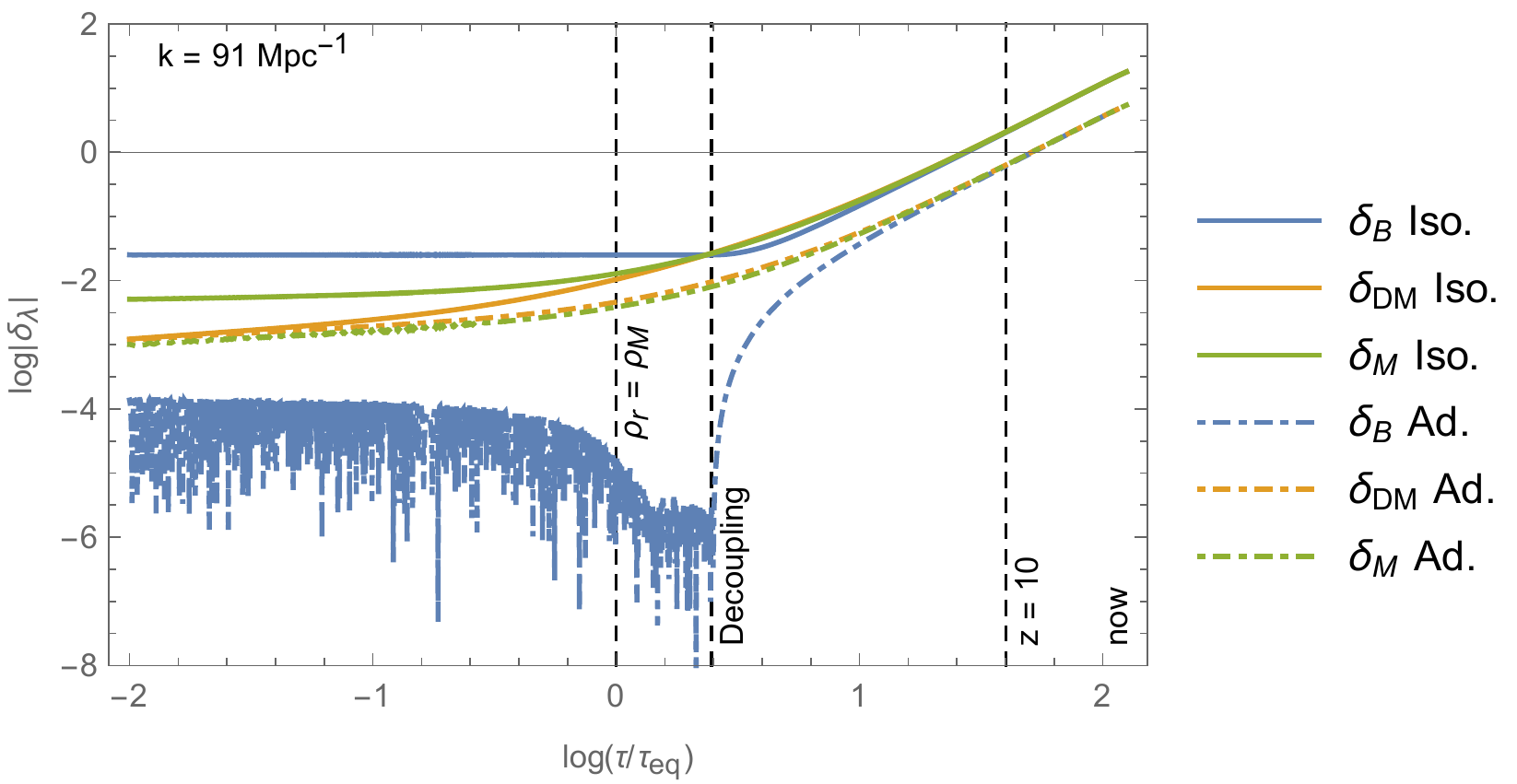}
	\caption{The evolution of baryon (blue), dark matter (orange), 
    and total matter (green) perturbations at $k = 1.4 k_s = 91\,\text{Mpc}^{-1}$ with $\delta_{B,\,0}$ determined by:
	1) (dashed lines) $\mathcal{R} = 4.7 \times 10^{-5}$, appropriate for a scenario with only primordial adiabatic perturbation, and 
	2) (solid lines) Including isocurvature perturbations; following equation \eqref{eq:dB_ini}, $\delta_B = 0.025$ for the $k = 91 \, \mathrm{Mpc}^{-1}$ mode if $k_s = 65 \, \mathrm{Mpc}^{-1}$.  This scenario is appropriate to the Higgs relaxation scenario considered in this work. 
	\label{fig:Isocurvature_mode_evolution}}
\end{figure}

An example of the evolution of a single mode is shown in Fig.~\ref{fig:Isocurvature_mode_evolution}.  We take $k_s = 65 \, \mathrm{Mpc}^{-1}$, and consider the mode at $k = 1.4 k_s = 91 \, \mathrm{Mpc}^{-1}$.  The baryon density contrast given by Eq.~\eqref{eq:dB_ini} is then 0.025.  The evolution of the baryon, dark matter, and total matter perturbations are shown with solid lines.  For completeness, we have also shown the evolution without the isocurvature modes in dashed lines (without accounting for Silk damping).  We see that as expected the isocurvature perturbation does not evolve until decoupling; afterwards, it grows.  Prior to decoupling, it enhances perturbations in dark matter and total matter.

\begin{figure}
	\includegraphics[width = 0.6\columnwidth]{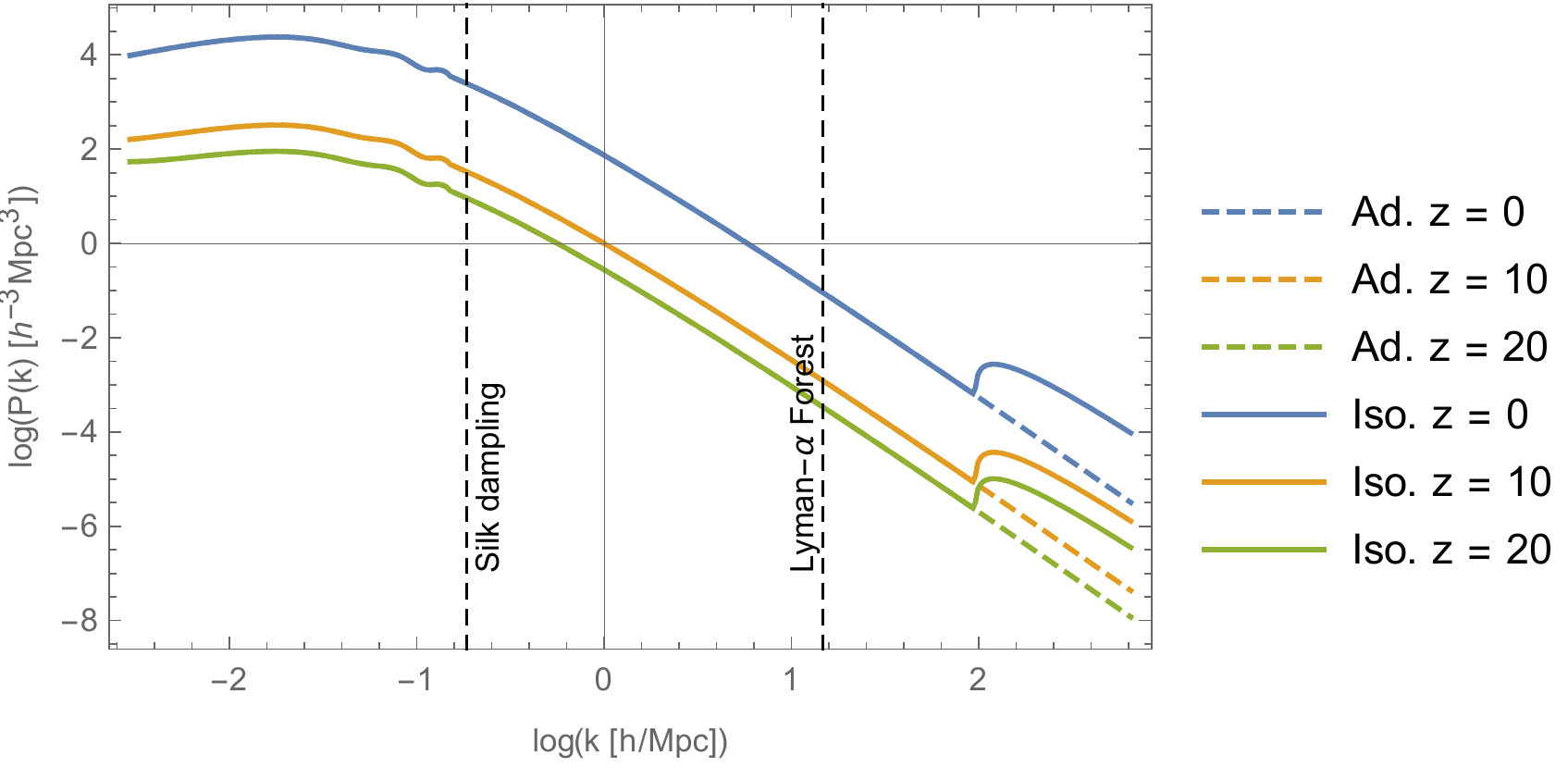}
	\caption{Total matter power spectra at $z = 0,\,10,\,\text{and}\,20$
	for the cases that the primordial perturbations are produced by 1)
	only inflaton (dashed line), and 2) inflaton plus relaxation leptogenesis
	with $k_{s} = 65\,\mathrm{Mpc}^{-1}$ and $N_{\mathrm{last}} = 46.5$.  Adiabatic perturbations to the right of the first dashed line are affected by Silk damping, although isocurvature contributions are not.  The power spectrum to the left of the second dashed line are constrained by the Lyman-$\alpha$ constraints. The bump on the right edge of the plot describes the contribution on the isocurvature perturbations.
    \label{fig:Power_Spectrum}}
\end{figure}

In Fig.~\ref{fig:Power_Spectrum}, we present the total matter
power spectrum, which is given by
\begin{equation}
P\left(k,\,z\right) = \frac{2\pi^{2}}{k^{3}}\mathcal{P}_{m}\left(k,\,z\right)
= \frac{2\pi^{2}}{k^{3}}\delta_{m}^{2}\left(k,\,z\right).
\end{equation}
By varying $\delta_{B,\,0}$, we have found that for $\delta_{B,\,0} \sim 0.025$, the total matter perturbation $\delta_{m}=\left(\Omega_{B}\delta_{B}+\Omega_{DM}\delta_{DM}\right)/\Omega_{m}$
reaches the non-linear regime ($\delta_{m}\gtrsim 1$) much earlier than it would if only the adiabatic fluctuation were present.  Thus, in the Higgs leptogenesis model, structure formation begins earlier, which allows for earlier star formation.  In the next section, we will use this modified history of structure formation to explain the cosmic infrared radiation excess.

\section{Isocurvature Perturbations and the Cosmic Infrared Background Observations}
\label{sec:does_iso_work}

In the above sections, we demonstrated that the Higgs relaxation leptogenesis scenario generates baryonic isocurvature perturbations and studied their evolution in the early universe.  Now, we proceed to connect the above results to the observed CIB radiation.  The isotropic flux (or absolute intensity) of the CIB is difficult to determine precisely due to the large uncertainty associated with the removal of the foreground signal, galactic components, and zodiacal light. Therefore, recent measurements concern the anisotropies (spatial fluctuation) of the CIB \cite{Helgason:2015ema}.  From these measurements, one can infer the isotropic flux from the power in the fluctuations of the CIB.

In section \ref{sec:intro}, we mentioned the currently unexplained excess in observations of anisotropies in the near-infrared cosmic radiation spectrum, $\delta F_{2-5\,\mu\text{m}}\left(5^{\prime}\right)\simeq0.09\,\text{nW}\,\text{m}^{-2}\text{sr}^{-1}$ at 5 arcmin between 2 and 5 $\mu\text{m}$  \cite{Kashlinsky:2014jja}.  This relative fluctuation entails that the amplitude of the power in the fluctuations is $F_{\mathrm{CIB}} \approx \delta F_{\mathrm{CIB}}/\Delta_{5^{\prime}}\sim1\,\text{nW m}^{-2}\text{sr}^{-1}$; one is then led to consider what sources could produce this radiation.  One possibility is faint galaxies; such an explanation is consistent with AKARI observations but not the Spitzer observations, due to the fact that Spitzer is able to resolve fainter point sources~\cite{Helgason:2016xoc}.
As discussed in Ref.~\cite{Kashlinsky:2014jja, Helgason:2015ema}, one possible source is early (population III) stars, at $ z \approx 10$.  Such stars, if they exist, will contribute significantly to the CIB and live only for a short cosmological time. 
In this case, the power in the fluctuations is equivalent to the isotropic flux due to the early stars~\cite{Helgason:2016xoc}.

However, Ref.~\cite{Kashlinsky:2014jja, Cooray:2012dx, Yue:2012dd} shows that in the typical model of structure formation, with only adiabatic perturbations, one requires either an abnormally large stellar formation efficiency and/or an abnormally large radiation efficiency to produce the requisite amount of CIB radiation.  We now demonstrate that the presence of isocurvature perturbations alters this conclusion.  In our model, the isocurvature perturbations produced by Higgs relaxation leptogenesis cause a larger percentage of the mass in the early universe to be in collapsed halos which evolve non-linearly and can support early star formation.  Therefore, the comparably large isotropic CIB flux (that is, the power in the fluctuations) can be produced with a reasonable values for the stellar formation efficiency and radiation efficiency.

As the above discussion outlines, we are interested in the isotropic CIB flux due to early stars.  The contribution from the first stars forming inside collapsed halos can be estimated by~\cite{Kashlinsky:2014jja} 
\begin{equation}
F_{FS}\simeq\frac{c}{4\pi}\epsilon\rho_{B}c^{2}f_{\mathrm{Halo}}f_{*}z_{\mathrm{eff}}^{-1}=9.1\times10^{5}\epsilon f_{\mathrm{Halo}}f_{*}\left(\frac{\Omega_{B}h^{2}}{0.0227}\right)\left(\frac{10}{z_{\mathrm{eff}}}\right)\,\text{nW}\,\text{m}^{-2}\text{sr}^{-1},
\end{equation}
where $f_{\mathrm{Halo}}$ is the mass-fraction of the universe inside
halos, $f_{*}$ is the star formation efficiency, $\epsilon$ is the
radiation efficiency, and $z_{\mathrm{eff}}$ is the effective redshift.  One then finds that the halo fraction at $z=10$ is given by
\begin{equation}
	f_{\mathrm{Halo}} = 0.16 \left(\frac{0.007}{\epsilon}\right) \left(\frac{10^{-3}}{f_{*}}\right)
\left(\frac{F_{FS}}{F_{\mathrm{CIB}}}\right) 
.
    \label{eq:f_halo_exp}
\end{equation}
In order to have $F_{FS} = F_\mathrm{CIB} = 1 \, \mathrm{nW} \; \mathrm{m}^{-2} \; \mathrm{sr}^{-1}$ (the value implied by the assumption that early stars explain the observed CIB anisotropy) for reasonable values of the parameters are $\epsilon \approx 0.007$ and $f_{*} \lesssim 10^{-3}$, one must have $f_\mathrm{Halo} \gtrsim 0.16$.  (The value of $\epsilon$ comes from the hydrogen burning phase of early stars, which are fully convective and radiate close to the Eddington limit (see Ref.~\cite{Kashlinsky:2016sdv}); our preferred value of $f_*$ comes from the same reference.) We fix $\epsilon$ and $f_*$ at their upper bounds and show that with isocurvature perturbations one can have $f_\mathrm{Halo} \approx 0.16$, which one cannot accomplish with only adiabatic perturbations.

To compute the fraction of matter in collapsed halos, we adopt the Press-Schechter formalism \cite{1974ApJ...187..425P}. An overdense region which in the linear theory would have present size $R$ has in fact collapsed and formed structure by the time when the average density contrast $\delta_{R}\left(\boldsymbol{x},\,t\right)$ exceeds $\delta_{c}\cong1.686$, as calculated in the linearized theory defined by equations \eqref{eq:linear_theory} above.  The average matter density contrast is computed by smoothing the spectrum  
\begin{equation}
	\delta_{R}\left(\boldsymbol{x},\,t\right) = \int d^{3}y \, \delta_{m}
	\left(\boldsymbol{x} + \boldsymbol{y},\,t\right) W_{R}\left(\boldsymbol{y}\right),
\end{equation}
where a window function $W_{R}\left(\boldsymbol{y}\right)$ is used to smooth the matter density so that one attains an average; we use the top-hat function
\begin{equation}
	W_{R}\left(\boldsymbol{y}\right) = \frac{3}{4\pi R^{3}}
    \theta\left(R-\left|\boldsymbol{y}\right|\right).
    \label{eq:top-hat-fn}
\end{equation}
which has the Fourier transform $W_{R}\left(k\right)=3j_{1}\left(kR\right)/kR$.   Using this window function, the mass contained in a sphere of radius $R$ is approximately
\begin{equation}
	M\left(R\right) = \frac{4}{3}\pi R^{3}\rho_{m,\,0},
    \label{eq:MandR}
\end{equation}
where $\rho_{m,\,0}$ is the present average matter density of the universe.
The smoothed matter density contrast $\delta_{R}\left(\boldsymbol{x},\,t\right)$ computed in this way is itself a Gaussian random field, whose variance $\sigma_{R}\left(t\right)$ is given by 
\begin{equation}
	\sigma_{R}^{2}\left(t\right)
    \equiv \left\langle \delta_{R}^{2}\left(\boldsymbol{x},\,t\right)\right\rangle
    = \int_{0}^{\infty}\frac{dk}{k}\mathcal{P}_{m}\left(k,\,t\right)\left|W_{R}\left(k\right)\right|^{2},
\label{eq:integrand}
\end{equation}
Using equation \eqref{eq:MandR}, one can solve for radius $R$ in terms of $M$, the total mass contained inside.  Substituting this into $\sigma_R(t)$ gives the variance $\sigma_M(t)= \sigma_{M(R)}(t)$ as a function of enclosed mass $M$.  

\begin{figure}
\includegraphics[scale=.7]{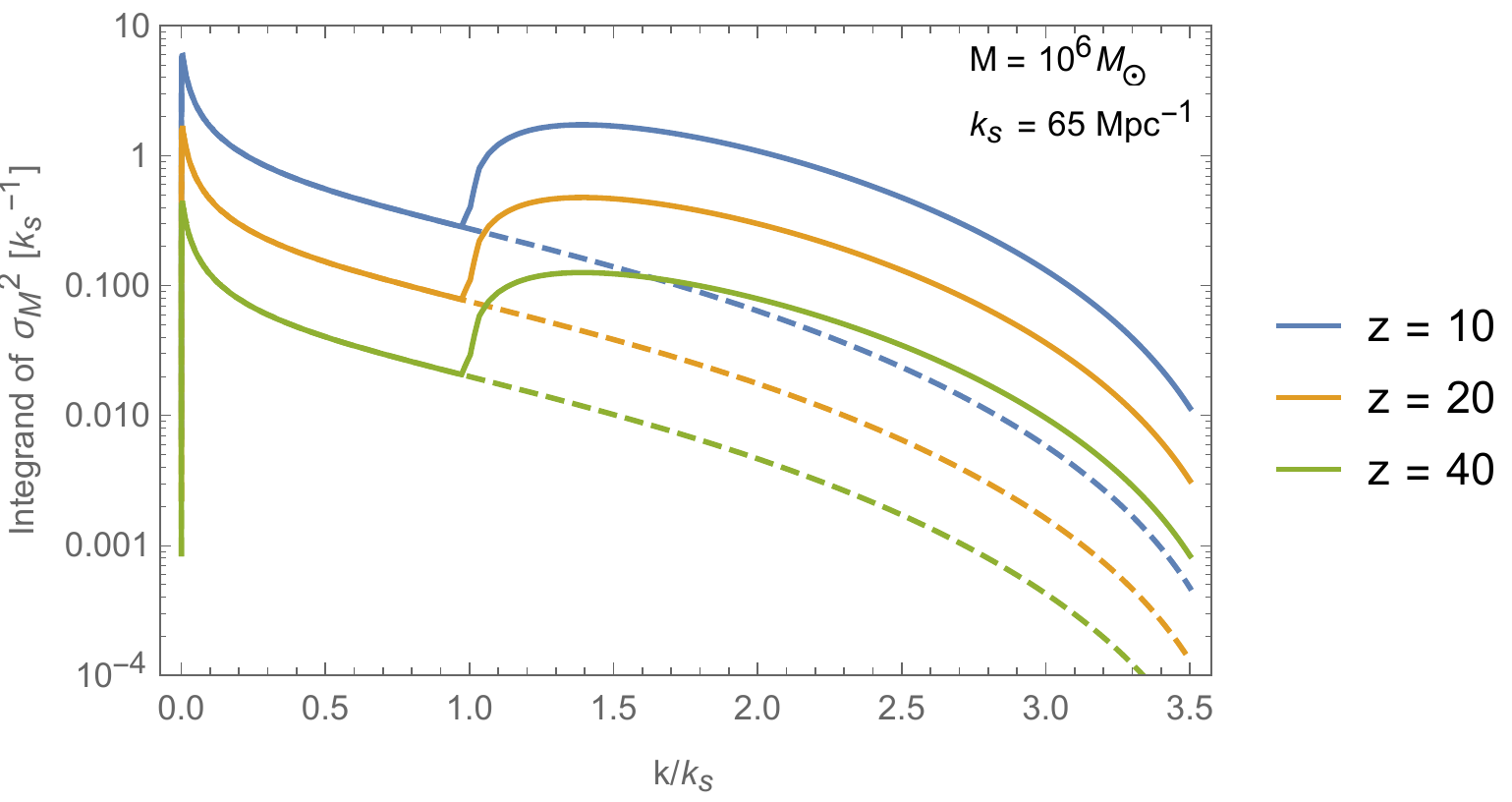}
\caption{The integrand for $\sigma_M^2$ as a function of $k \slash k_s$ for various redshifts $z$. The dashed lines are for the power spectrum with only adiabatic modes. The solid lines are for the power spectrum with isocurvature perturbation turned on at $k_s = 65\,\mathrm{Mpc}^{-1}$. We see that the integrand  is dominated by $k \slash k_s \cong 1.4$ for the isocurvature case, neglecting the peak near zero which is present near for both the scenario with and without the isocurvature contribution.}
\label{fig:sigma_integrand}
\end{figure}

The integrand of Eq.~\eqref{eq:integrand} is shown in Fig.~\ref{fig:sigma_integrand}, where we have fixed the radius to correspond to a mass of $10^6 M_\odot$.  This figure shows that for multiple redshifts, the integrand is peaked at $k \cong 1.4 k_s$.  This justifies the claim that our signal is dominated by the contribution in this region, which was mentioned above and which motivated our choice of $k \slash k_s = 1.4$ as a reference point for characterizing $\delta_B$.

\begin{figure}
\includegraphics[height=0.16\paperheight]{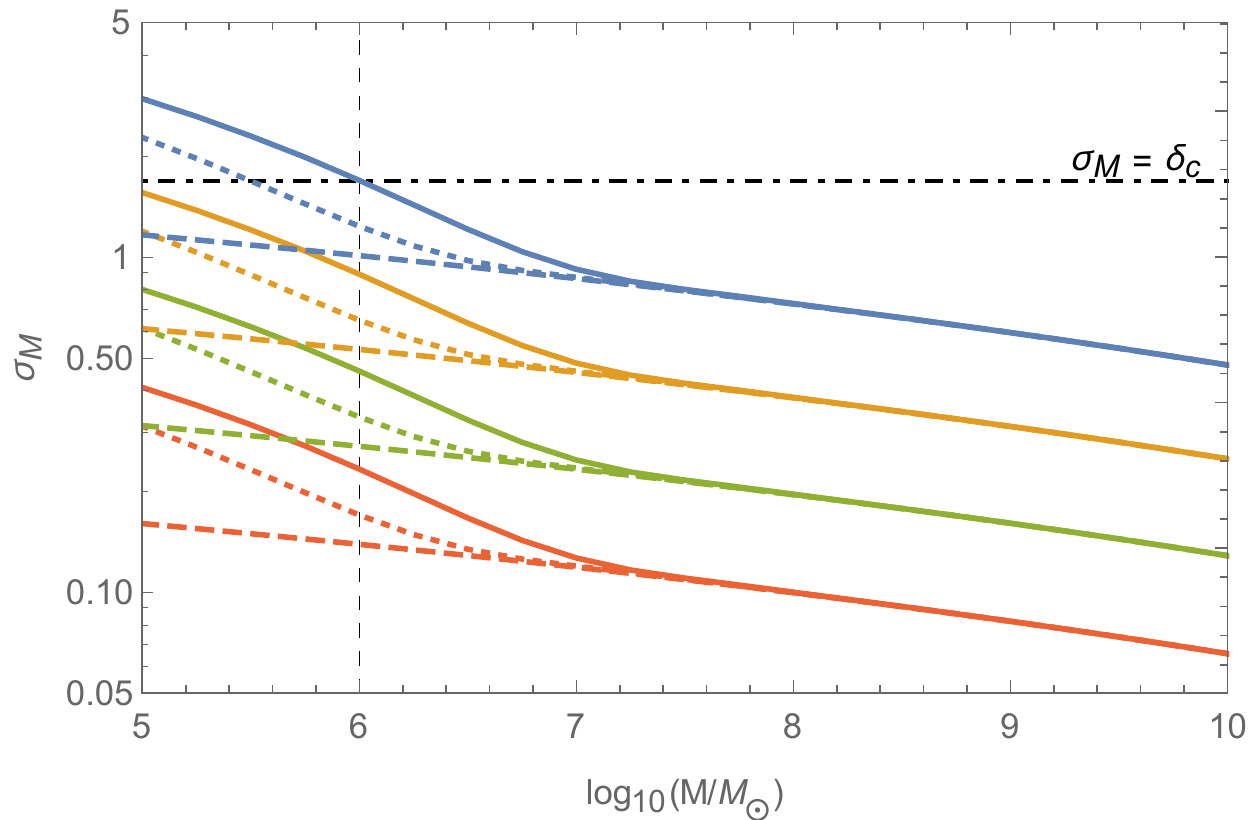}
\includegraphics[height=0.16\paperheight]{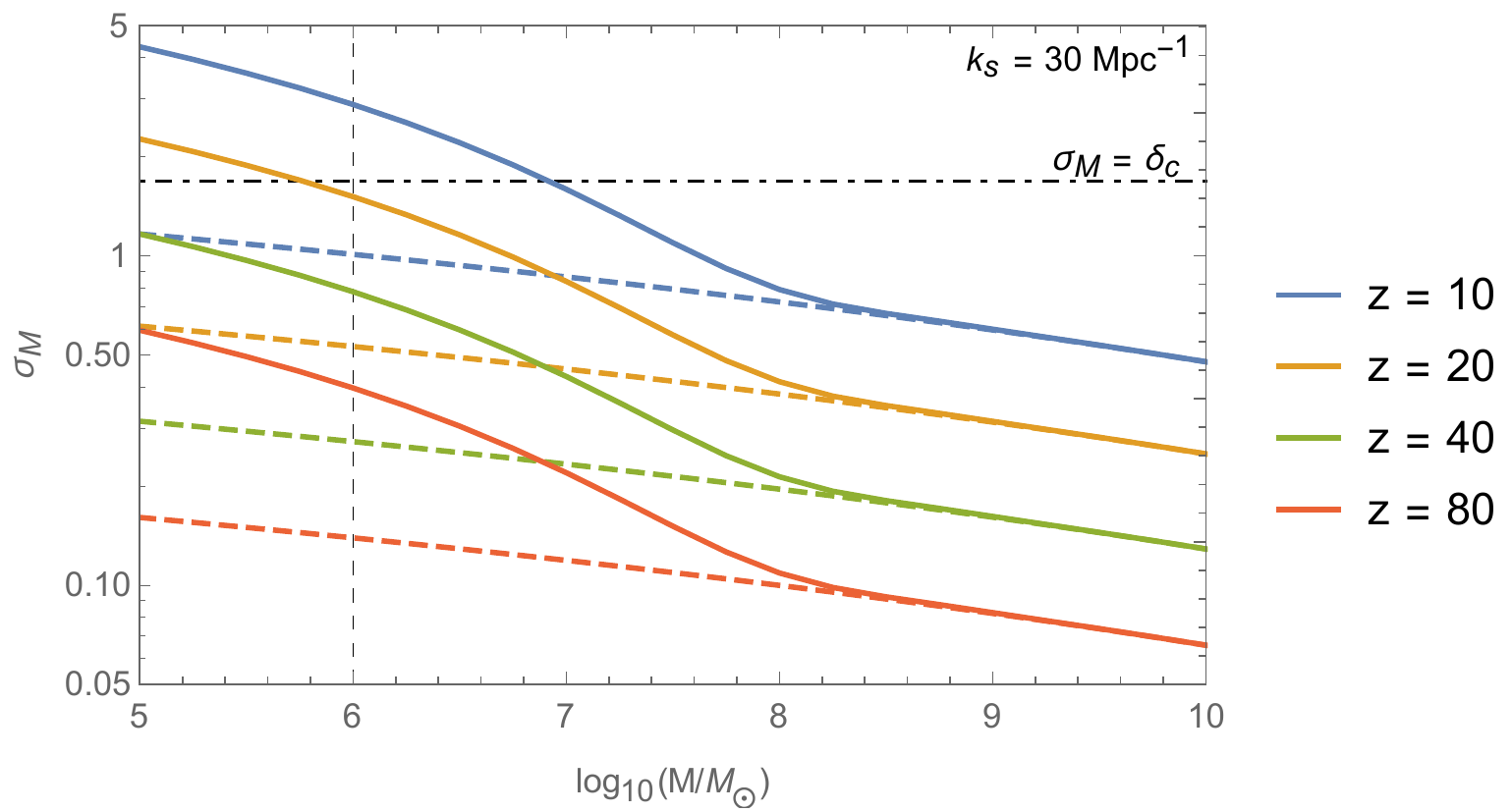}

\caption{The variance $\sigma_{M}$ of the smoothed density contrast at various redshifts and mass scales.  Left: the dashed lines show the results with only adiabatic modes ($\mathcal{R}=\sqrt{A_{S}}\protect\cong4.7\times10^{-5}$), while the solid (dotted) lines also includes isocurvature perturbations for $k \ge 65\,\mathrm{Mpc}^{-1}$ ($100\,\mathrm{Mpc}^{-1}$), with $\delta_{B,\,0} = 0.025$ at $k = 1.4 k_s$. Right: The solid lines correspond to $k_s = 30 \, \mathrm{Mpc}^{-1}$.  The black dash-dotted horizontal lines denotes the value $\sigma_{M}=\delta_{c}$; structure formation occurs above this line. \label{fig:sigma_M_2}}
\end{figure}

Fig.~\ref{fig:sigma_M_2} shows this $\sigma_{M\left(R\right)}$ at various mass scales and redshifts.  On both plots, the dashed lines show $\sigma_{M(R)}$ including only adiabatic perturbations, while the solid and dotted lines includes the isocurvature perturbations generated by the Higgs relaxation mechanism, which we emphasize only exist for $k\ge k_s$.  The plot of the left shows the results for $k_s = 65 \, \mathrm{Mpc}^{-1}$ (solid) and $k_s = 100 \, \mathrm{Mpc}^{-1}$ (dotted); on the right, we show the results for $k_s = 30 \, \mathrm{Mpc}^{-1}$.  As expected, we see that $k_s = 65 \, \mathrm{Mpc}^{-1}$ leads to a larger deviation from the adiabatic-only model than $k_s = 100 \, \mathrm{Mpc}^{-1}$.  (We have used the initial value of $\delta_{B,\,0}$ and $\mathcal{R}$ given in the sections above.)  On both plots, the black dash-dotted horizontal line corresponds to the critical variance; above this, a significant portion of the halos of a particular mass evolve non-linearly.    

Focusing on the $k_s = 65 \, \mathrm{Mpc}^{-1}$ solid lines (left), we see that halos of mass $10^5 M_\odot$ would collapse around $z=20$ while those of mass $10^6 M_\odot$ would collapse around $z = 10$ in the Higgs relaxation model; this contrasts to the standard picture, in which such halos would form later.  At any given $z$, there are more halos with mass $M \lesssim 10^7 M_\odot$ in the Higgs relaxation scenario than in the typical scenario which has only adiabatic perturbations.  Because the density contrast at mass scales $M\gtrsim10^{7}M_{\odot}$ is unaffected, the observed large scale structure is unchanged. 

For the $k_s = 100 \, \mathrm{Mpc}^{-1}$ (dotted) lines, the formation of small halos is still enhanced with respect to the adiabatic-perturbations only scenario; however, these halos form later.  
We focus on $10^6 M_\odot$ because such halos are near the lower bound of halos that can efficiently support star formation through molecular hydrogen cooling~\cite{Abel:1996kh, Tegmark:1996yt, Abel:2001pr, Bromm:2001bi, Bromm:2009uk}.  Production of these $10^6 M_\odot$ halos is not significantly enhanced for $k_s = 100 \, \mathrm{Mpc}^{-1}$, which means that we require $k_s \lesssim 100 \, \mathrm{Mpc}^{-1}$ to explain the CIB.  The plot on the right shows the situation with $k_s = 30 \, \mathrm{Mpc}^{-1}$; we see that halos of mass $10^6 M_\odot$ form earlier, around $z = 20$. We see that increasing $k_s$ would bring us into conflict with optical depth measurements.  Therefore, to explain the CIB excess, we require $k_s \approx 65 \, \mathrm{Mpc}^{-1}$.

We now show that we make sufficiently many collapsed halos.  Using the variance $\sigma_M$ in the matter density contrast, we calculate the probability that a region with mass $M\left(R\right)$ has an average density contrast $\delta_R$ exceeding $\delta_{c}$ at redshift $z$ \cite{1974ApJ...187..425P}, which is
\begin{equation}
	f_{\mathrm{Halo}}\left(M,\,z\right) = P\left(\delta_{R\left(M\right)}>\delta_{c}\right)
    = \frac{1}{2}\left[1-\mathrm{erf}\left(\frac{\delta_{c}}{\sqrt{2}\sigma_{M}\left(z\right)}\right)\right].
\end{equation}
This is equivalent to the fraction of mass which is collapsed halos of mass $M$, as smaller structures form earlier.

\begin{figure}
	\includegraphics[height=0.18\paperheight]{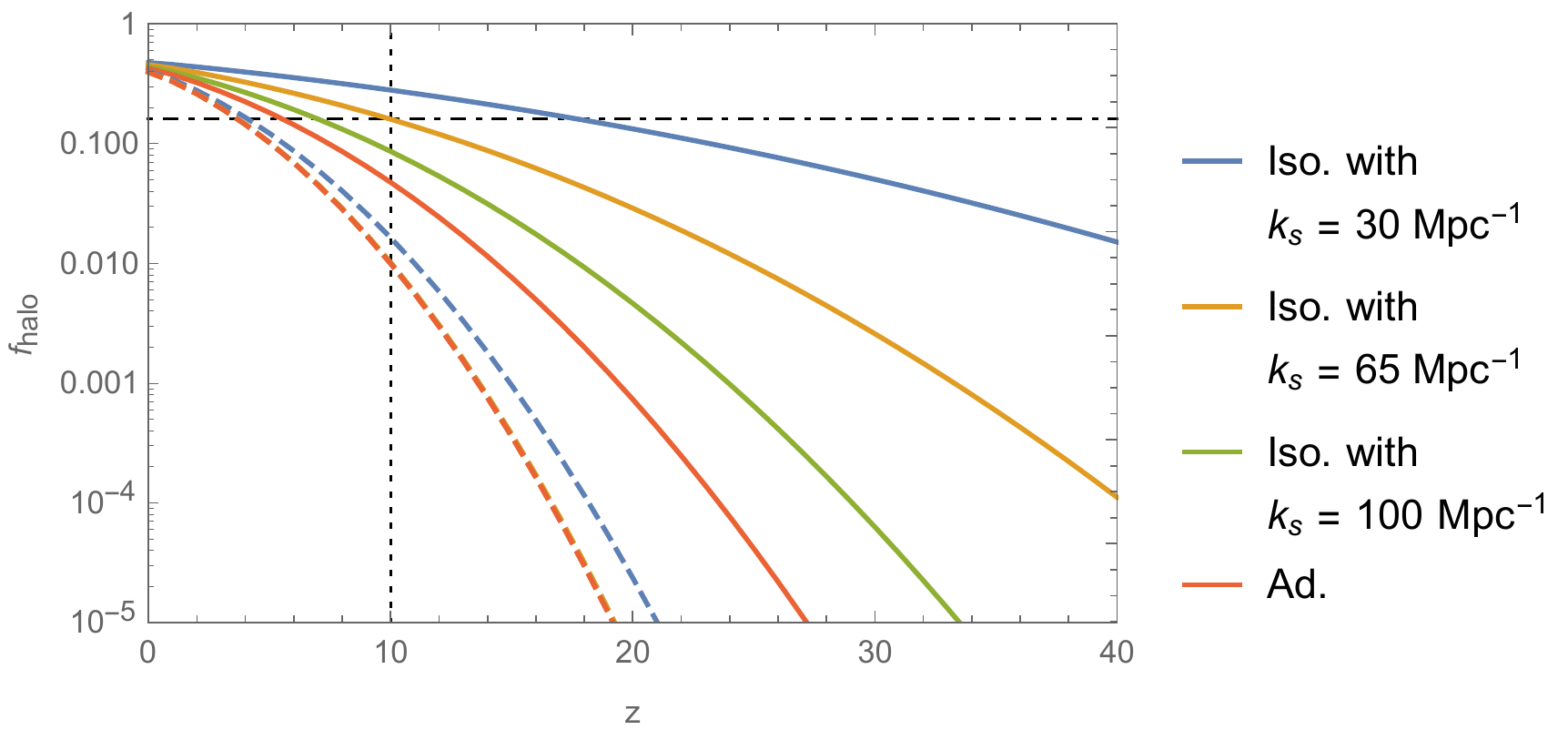}
	\caption{$f_{\mathrm{Halo}}\left(M,\,z\right)$, the mass fraction of the universe inside the collapsed halos of mass $M$, evaluated for $M=10^{6}M_{\odot}$ (solid lines) and $10^{8}M_{\odot}$ (dashed lines), as a function of redshift $z$.  Red lines represent the scenario with only adiabatic perturbations, while blue, yellow, and green lines represent the scenario which includes isocurvature perturbations for  $k \ge k_s = 30$, $65$, and $100\,\mathrm{Mpc}^{-1}$, respectively.  The vertical dotted black line emphasizes $z=10$; early stars at this redshift can potential explain the CIB excess.  The horizontal dot-dashed black line indicated $f_\mathrm{Halo} = 0.16$; as discussed in the text, a model explains the CIB observations for reasonable $\epsilon$ and $f_*$ values only if $f_\mathrm{Halo}$ takes this value for star-forming halos. ($\mathcal{R}$ and $\delta_{B,\,0}$ take the same values as in Fig.~\ref{fig:sigma_M_2}.)}
    \label{fig:f_halo}
\end{figure}

The results of this calculation is presented in Fig.~\ref{fig:f_halo} for $M = 10^6 M_\odot$ (solid lines) and $M = 10^8 M_\odot$ (dashed lines), first with only adiabatic modes (red lines) and then including the isocurvature modes (blue, yellow, and green lines).  The Higgs relaxation scenario, with the isocurvature modes, is more efficient in halo formation; however, as expected from Fig.~\ref{fig:sigma_M_2}, the gain in efficiency is more pronounced for smaller halos.  The vertical dashed line denotes $z=10$; early stars at this time contribute significantly to the CIB.  Therefore, we desire that halos large enough to support star formation ($\gtrsim 10^6 M_\odot$) have formed by this redshift.

As explained above, we will have sufficient stars to produce the inferred CIB excess for reasonable values of the radiation efficiency $\epsilon$ and star formation efficiency $f_*$ if $f_\mathrm{halo} \approx 0.16$ for halos large enough to support star formation.  Therefore, we have included a horizontal black dot-dashed line at $f_\mathrm{halo} = 0.16$.  In the scenario calculated with the Higgs relaxation isocurvature perturbations, the $10^6 M_\odot$ line indeed passes near $f_\mathrm{halo} = 0.16$ at $z=10$ if we take $k_s = 65 \, \mathrm{Mpc}^{-1}$ (yellow).  As expected from the above discussion, $k_s = 30 \, \mathrm{Mpc}^{-1}$ (blue) results in a larger percentage of the mass in collapsed halos and $k_s = 100 \, \mathrm{Mpc}^{-1}$ (green) a smaller percentage.  In the scenario which includes only adiabatic perturbations, the $f_\mathrm{Halo}$ line for $10^6 M_\odot$ is significantly suppressed; this is the source of the claim that unreasonably large radiation efficiency or star formation efficiency is required in the standard picture.  We see that for $k_s \approx 65 \, \mathrm{Mpc}^{-1}$ a sufficiently large percentage of the mass is in halos $\sim 10^6 M_\odot$ to account for the inferred contribution from early stars to the isotropic CIB flux.

\begin{figure}
\includegraphics[scale=.8]{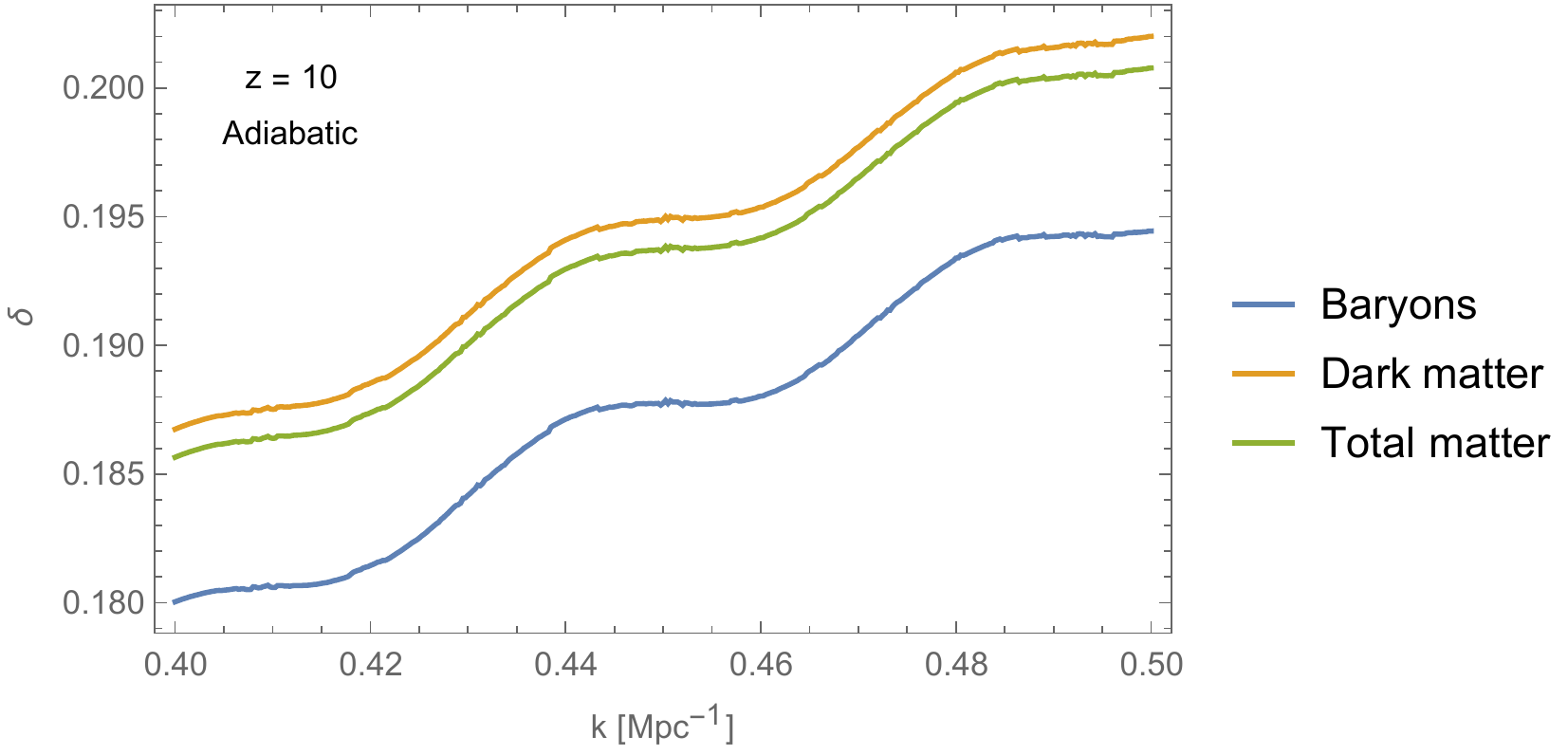}
\caption{The density contrast at scales $k = 0.40 \; \mathrm{Mpc}^{-1}$ to $k = 0.50 \; \mathrm{Mpc}^{-1}$, for which only the adiabatic perturbations contribute.  The lines show the baryonic, dark matter, and total matter perturbations.  $k \sim 0.45 \; \mathrm{Mpc}^{-1}$ corresponds to the 5 arcminute scale probed observationally. 
\label{fig:delta_at_5_arcminutes}
}
\end{figure}

Finally, we note that the isotropic CIB flux from early stars is inferred from the anisotropic flux measured at scales of 5 arcminutes, corresponding to $k \sim 0.45 \; \mathrm{Mpc}^{-1}$, which is much smaller than $k_s = 65 \; \mathrm{Mpc}^{-1}$.  Therefore, only adiabatic modes contribute at this scale; the density contrast is shown in Fig.~\ref{fig:delta_at_5_arcminutes}.  We see that at $k \sim 0.45 \; \mathrm{Mpc}^{-1}$ the density contrast is $\sim 10 \%$, consistent with the calculations in \cite{Kashlinsky:2014jja} and for similar reasons, consistent with the observational anisotropic data. (Note that although there is a difference between two-dimensional and three-dimensional power spectra, the difference should be order 1.) Therefore, the isocurvature perturbations considered here explain the inferred contribution of the early stars to the isotropic CIB excess without overproducing an anisotropic contribution.

To summarize, in our model, structure is generated by adiabatic perturbations at the large scale and the isocurvature perturbations at smaller scales.  The isocurvature perturbations are responsible for causing more halos ($10^6 M_\odot$) to evolve non-linearly, and hence, we make a sufficient number of stars to explain the isotropic CIB radiation inferred from the anisotropic measurements without a large stellar formation efficiency.  However, these halos are distributed in accordance with the larger-scale adiabatic perturbations, and the scale of the CIB anisotropy is accounted for by this larger-scale structure.  This provides an elegant solution as to the source of the observed CIB radiation fluctuations.

We also remark that in general, the early creation of population III stars is constrained by the optical depth measurements of the CMB.  We note that recent analyses of the Planck 2015 optical depth data in fact prefers early star formation, particularly if one includes self-regulated population III stars~\cite{Miranda:2016trf}.  If star formation occurred much earlier than $z = 10$, as for $k_s \lesssim 30 \, \mathrm{Mpc}^{-1}$, then this scenario would conflict with optical depth measurements.  However, as noted, for $k_s = 65 \, \mathrm{Mpc}^{-1}$ the star formation occurs around $z=10$.

\section{Available Parameter Space}
\label{sec:parameter_space}

\begin{figure}
\includegraphics[scale=.6]{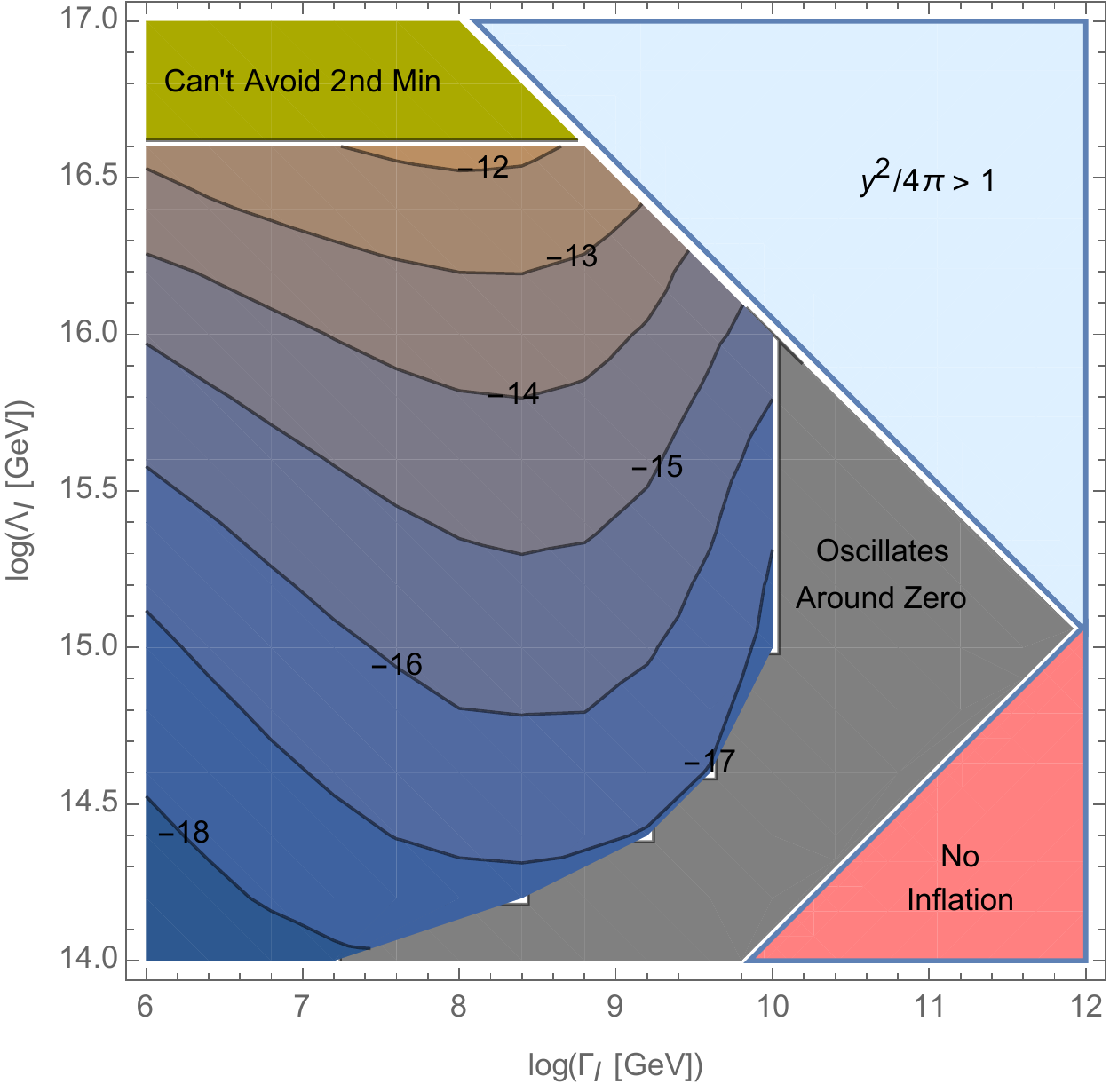}
\caption{The lepton-number-to-entropy ratio, $Y$, from Higgs relaxation leptogenesis, with $N_\mathrm{last}$ set in order to explain the CIB excess and with the effective operator \eqref{eq:O6operator} generator by thermal loops, so that $\Lambda_n \sim T$.  In the upper left corner, the $N_\mathrm{last}$ required to explain the CIB excess is such that the Higgs VEV probes the second minimum in the Higgs potential.  In the lower right, $\Gamma_I > 3H_I$ and inflation is not successful.  In the gray region, washout is sufficiently large to cause the lepton asymmetry to oscillate around zero at the end of our numerical analysis; the final value will be quite small.  As in Ref.~\cite{Yang:2015ida,Gertov:2016uzs}, the neutrino Yukawa coupling is chosen to suppress thermal leptogenesis; in the upper right of the plot, this condition leads into the non-perturbative regime.  We see that there is no parameter space in which a sufficiently large asymmetry is generated.}
\label{fig:CIB_T_asym}
\end{figure}

In this section, we present plots of the parameter space in which Higgs relaxation leptogenesis can both explain the observed matter-antimatter asymmetry of the universe and the observations of the cosmic infrared background radiation.  We note that Higgs relaxation leptogenesis is only one potential source of baryonic isocurvature perturbations; other sources include curvaton models (proposed in~\cite{Enqvist:2001zp,Lyth:2001nq,Moroi:2001ct}; see also 
\cite{Postma:2002et,Postma:2003gc,Postma:2004an,Harigaya:2014bsa}) and warm inflation (e.g.,~\cite{BasteroGil:2011cx,Bastero-Gil:2014oga}).  In general, any model which produces baryonic isocurvature perturbations similar to those discussed above can account for the observed CIB excess.

In these plots, we choose $N_\mathrm{last}$, $\Lambda_I$, and $\Gamma_I$ such that $k_s = 65 \; \mathrm{Mpc}^{-1}$; then we determine the initial vacuum expectation value of the Higgs field using equation \eqref{eq:real_phi0}, which includes the curvature of the Higgs potential.  As discussed in section \ref{sec:Primordial-Baryonic-Isocurvature}, and shown explicitly in Fig.~\ref{fig:delta_B_parameter_space}, this leads to $\delta_{B}(k\slash k_s = 1.4) \approx 0.025$ throughout parameter space, sufficient to explain the CIB observations.  (Regions where the requisite initial VEV probes the second vacuum in the Higgs potential are denoted on the plots.)  We note that we include one-loop corrections to the Higgs potential and two-loop corrections to the running couplings; for details regarding the potential (including the specific values for the Higgs mass and top quark mass used), please see the Higgs relaxation leptogenesis analysis in~\cite{Yang:2015ida}.

As discussed in~\cite{Kusenko:2014lra,Pearce:2015nga,Yang:2015ida}, there are several different mechanisms of generating the $\mathcal{O}_6$ operator; one can use thermal loops, leading the scale $\Lambda_n \sim T$, or one can introduce heavy fermions, leading to a scale $\Lambda_n \sim M_n$, a constant.  The parameter space for these two options was explored extensively in Ref.~\cite{Yang:2015ida}, with the result that when the initial Higgs vacuum expectation value was set by quantum fluctuations, the largest lepton-asymmetry-to-entropy ratio that was possible with $\Lambda_n \sim T$ was $Y \sim 10^{-12}$ (Fig.~12 of Ref.~\cite{Yang:2015ida}), while for $\Lambda_n \sim M_n$, parameter space was available, but in the regime in which the use of effective field theory to describe the $\mathcal{O}_6$ operator was questionable.

We mentioned above that in Ref.~\cite{Kusenko:2014lra,Yang:2015ida,Gertov:2016uzs}, we took $N_\mathrm{last} = 8$ out of an abundance of caution to avoid baryonic isocurvature constraints, but the actual limit is much weaker.  Here, $N_\mathrm{last}$ is set by \eqref{eq:N_last_limit}, which is generally larger ($N_\mathrm{last} \sim 40$ typically).  As explained in Appendix \ref{ap:analytical}, the final asymmetry is proportional to $\phi_0^2$, which grows as $N_\mathrm{last}$ in the limit of a flat potential.  (However, in our numerical analysis, we use equation \eqref{eq:real_phi0} which accounts for the curvature of the potential.)  Therefore, we expect the asymmetry to be enhanced as compared to our previous analysis, although not significantly.

This is illustrated for the $\Lambda_n = T$ case in Fig.~\ref{fig:CIB_T_asym}.  This figure shows contours of the lepton asymmetry to entropy ratio $Y$; regions with $Y \gtrsim 10^{-9}$ can account for the observed baryonic matter-antimatter asymmetry of the universe.  (We note that the original lepton asymmetry is redistributed between leptons and baryons by sphalerons.)  As compared to Fig.~12 of Ref.~\cite{Yang:2015ida}, the asymmetry is enhanced by about a little less than an order of magnitude; however, this is not sufficient to ensure a region of parameter space in which both a sufficiently large asymmetry is generated and the CIB excess is explained.

\begin{figure}
\includegraphics[scale=.6]{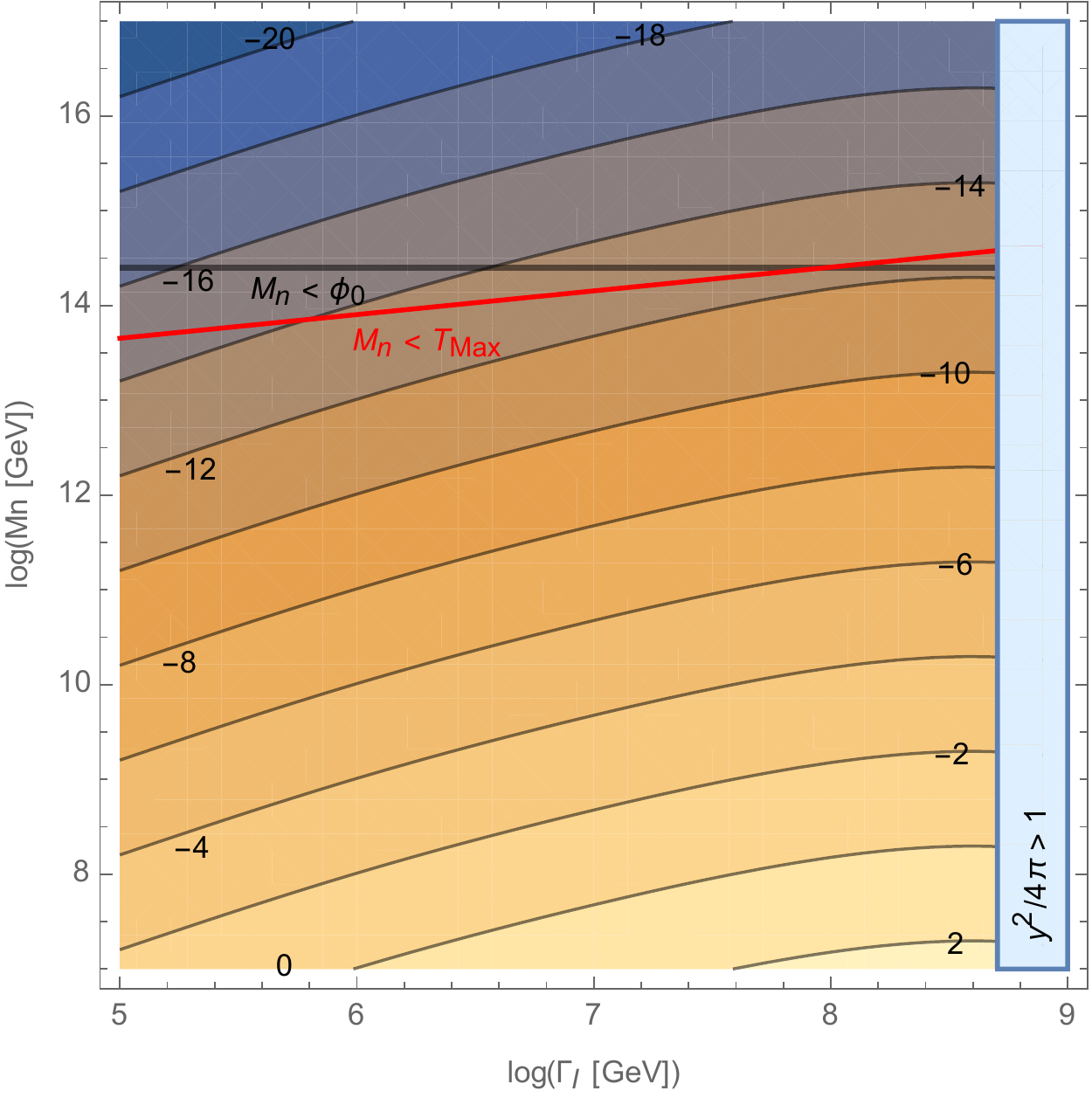}
\includegraphics[scale=.6]{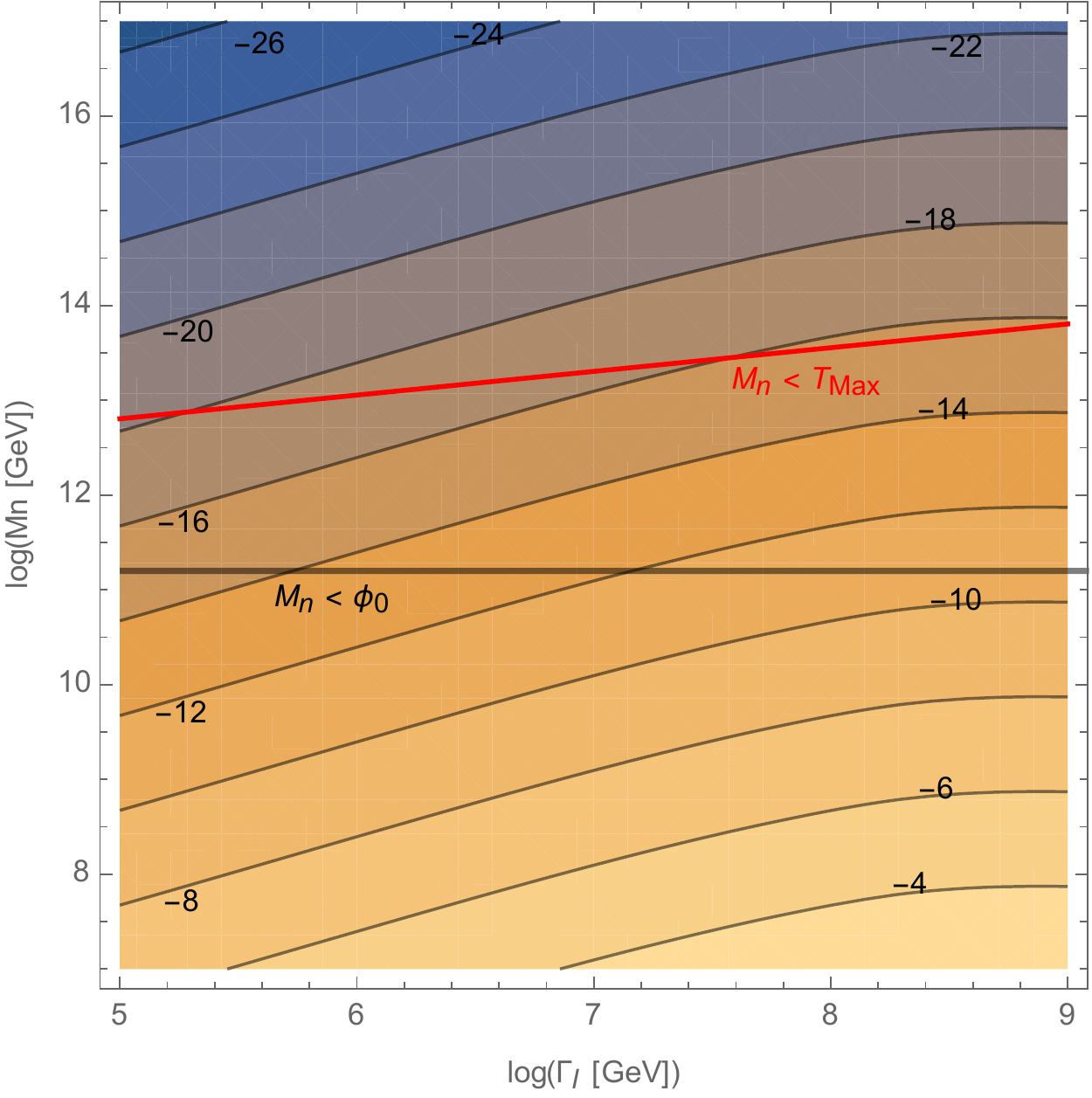}
\caption{The lepton-number-to-entropy ratio, $Y$, from Higgs relaxation leptogenesis, with $N_\mathrm{last}$ set in order to explain the CIB observations and with the effective operator \eqref{eq:O6operator} generator by new massive particles, so that $\Lambda_n \sim M_n$.  For these plots, $\Lambda_I = 5 \times 10^{16} \, \mathrm{GeV}$ (left) and $\Lambda_I = 10^{15} \, \mathrm{GeV}$ (right).   As in Ref.~\cite{Yang:2015ida,Gertov:2016uzs}, the neutrino Yukawa coupling is chosen to suppress thermal leptogenesis; for sufficiently large $\Gamma_I$, this condition leads into the non-perturbative regime.   We see that as sufficiently large asymmetry is generated, but in the regime in which the use of effective field theory with the operator \eqref{eq:O6operator} is questionable.}
\label{fig:CIB_Mn_asym}
\end{figure}

Therefore, we turn our attention to Fig.~\ref{fig:CIB_Mn_asym}, which instead has $\Lambda_n = M_n$, a constant.  We see that a sufficiently large asymmetry is generated for a wide range of inflaton couplings $\Gamma_I$ provided that the scale $M_n$ is small enough; the upper bound on $M_n$ becomes stronger as the inflation scale $\Lambda_I$ decreases.  Decreasing $\Lambda_I$ decreases the asymmetry, if $M_n$ and $\Gamma_I$ are held constant.

The red and gray lines illustrate where $M_n$, the scale in the $\mathcal{O}(6)$ effective operator, becomes less than $\phi_0$ and $T_\mathrm{max}$ respectively.  Below these lines, the use of effective field theory for $\mathcal{O}(6)$ is somewhat questionable.  This is not surprising as the same remark applied to the parameter space plots presented in~\cite{Yang:2015ida,Gertov:2016uzs}.  As discussed in~\cite{Yang:2015ida}, although the effective field theory description is questionable, we use it as an approximation as what would be found if an exact calculation in some UV-complete theory were done.  It was also shown in Ref.~\cite{Gertov:2016uzs} that this can be avoided in models with an extended scalar sector.

Subject to this caveat regarding the effective theory, we conclude that Higgs relaxation leptogenesis can successfully generate the observed matter-antimatter asymmetry while also generating isocurvature perturbations which enhance early star formation, explaining the observed CIB excess.  Thus, Higgs relaxation leptogenesis is a promising source for the desired baryonic isocurvature perturbations.

\section{Conclusion}

In this work, we have demonstrated that baryonic isocurvature perturbations at very small scales can cause halos of mass $10^6 M_\odot$ to collapse earlier than they would in the typical model of structure formation, which includes only adiabatic perturbations from inflation.  Since these halos can support the formation of population III stars, this leads to enhanced star formation in the early universe.  Therefore, the power in the fluctuations of the cosmic infrared background radiation measured by the Spitzer and AKARI space telescopes can be explained without invoking unreasonably large stellar formation efficiency or radiation efficiency.

As a source for these perturbations, we have used the Higgs relaxation leptogenesis model, in which the matter-antimatter asymmetry is produced via lepton-number-violating interactions in a plasma influenced by a time-dependent chemical potential produced by the relaxing Higgs vacuum expectation value.  If the initial vacuum expectation value of the Higgs field is set by quantum fluctuations, it will vary in different Hubble volumes, giving rise to slightly different baryon asymmetries.  These are the desired isocurvature perturbations.  The scale of these perturbations is set by number of $e$-folds the Higgs VEV grows through; we determined that we can explain the CIB observations if isocurvature perturbations exist for $k \gtrsim 65 \; \mathrm{Mpc}^{-1}$.  Finally, we illustrated the parameter space in which the Higgs relaxation model gives both successful leptogenesis and explains the CIB observations.

\textbf{Acknowledgements}

We thank P.~Adshead, C.Q.~Geng, N.~Gnedin, T.~Goto, and Y.~Tada for helpful discussions.  This work of A.K. and L.Y. was supported by the U.S. Department of Energy Grant No. DE-SC0009937.  M.K. and A.K. were also supported by the World Premier International Research Center Initiative (WPI), MEXT, Japan.  The work of L.P. was supported in part by the U.S. Department of Energy Grant No.  DE-SC0011842 at the University of Minnesota. L.Y. thanks National Center for Theoretical Sciences, Taiwan, for hospitality. The work of M.K. was supported by Grant-in-Aid for Scientific Research from the Ministry of Education, Science, Sports, and Culture (MEXT), Japan, No. 15H05889 and No. 25400248.

\appendix
\section{Relationship Between Lepton Number Density And Initial Higgs VEV}
\label{ap:analytical}

Within the Higgs relaxation leptogenesis paradigm, the generation of the asymmetry can occur through several mechanisms, even when the lepton-number-violating operator appears in the neutrino sector due to heavy right-handed Majorana states.  The asymmetry can be generated through particle production from the condensate as described by the Bogoliubov transformations~\cite{Pearce:2015nga}, or via lepton-number-violating scatterings occurring in the plasma, e.g., ~\cite{Kusenko:2014lra,Yang:2015ida}.  In this work, we are interested in the latter scenario, which requires a rapid production of plasma, perhaps even via some preheating mechanism.  This in turn entails that the thermal corrections to the Higgs potential, $\sim T^2 \phi^2$, tend to be large.  

In this case, the Higgs VEV relaxes rather rapidly, and throughout all of the parameter space shown in Figures \ref{fig:CIB_T_asym} and \ref{fig:CIB_Mn_asym}, the relaxation time scale is faster than the reheat time scale, determined by the decay rate of the inflaton.  This raises the concern that relaxation may proceed faster than the thermalization of the plasma, and therefore, that the finite temperature corrections to the Higgs potential are unreliable during relaxation.

According to Ref.~\cite{Harigaya:2013vwa}, the thermalization time scale is
\begin{equation}
t_\mathrm{th} \approx \alpha^{-16 \slash 5} \dfrac{m_I^{4 \slash 5}}{M_{Pl}^{3 \slash 5} \Gamma_I^{6 \slash 5}},
\end{equation}
where $m_I$ is the mass of the inflaton field, which is thus far undetermined in the Higgs relaxation scenario.  We note that for successful reheating, the inflaton must have available decay channels, despite the relatively large Higgs VEV $\phi_0$.  However, even at values $m_I \sim 10^{-5} \phi_0$ the inflaton is able to efficiently decay into electrons.  We have verified that in this limit, the thermalization time scale is faster than the relaxation time scale (using $\alpha \approx 1 \slash 40$ for the coupling, which accounts for its running at high scales).  Thus, it is consistent to consider the regime in which the relaxation time scale is less than the reheat time scale, $t_\mathrm{rlx} < t_{RH}$, and also that the Higgs potential during relaxation is dominated by the $T^2 \phi^2$ thermal correction.\footnote{We note that for $\phi_0 \gg m_I$, the Higgs bosons that participate in the scattering $h^0 \nu \leftrightarrow h^0 \bar{\nu}$ are produced via the thermalization of the plasma.  We also emphasize that we ensure that throughout the relaxation period, the energy density in the inflaton and produced radiation is greater than the energy density in the Higgs condensate.} 

Therefore, we here consider only the case that the effective potential of the scalar field is dominated by the thermal mass term 
\begin{equation}
V\left(\phi,T\right)=\frac{1}{2}\alpha_T^{2}T^{2}\phi^{2}.
\end{equation}
For the Standard Model Higgs field, the coefficient is $\alpha_T\approx\sqrt{\left(\lambda+\frac{9}{4}g^{2}+\frac{3}{4}g'^{2}+3h^{2}\right)/12}\approx0.33$ at the energy scale $\mu\approx10^{13}\,\text{GeV}.$ During the epoch of coherent oscillations of the inflaton, the energy density of the radiation as a function of time can be described by 
\begin{equation}
\rho_{r}\left(t\right)=\frac{m_{pl}^{2}\Gamma_{I}}{10\pi\left(t+t_{\mathrm{osc}}\right)}\left[1-\left(\frac{t_{\mathrm{osc}}}{t+t_{\mathrm{osc}}}\right)^{5/3}\right],
\end{equation}
where $t_{\mathrm{osc}}=\frac{2}{3}\sqrt{\frac{3}{8\pi}}m_{pl}/\Lambda_{I}^{2}$
and $\Gamma_{I}$ is the decay rate of the inflaton.  At all times we use an effective temperature for the plasma given by $\rho_r = \pi^2 g_* T^4 \slash 30$; as discussed, this is valid for $t > t_{th}$.

\begin{figure}
\centering{}\includegraphics[width=0.6\paperwidth]{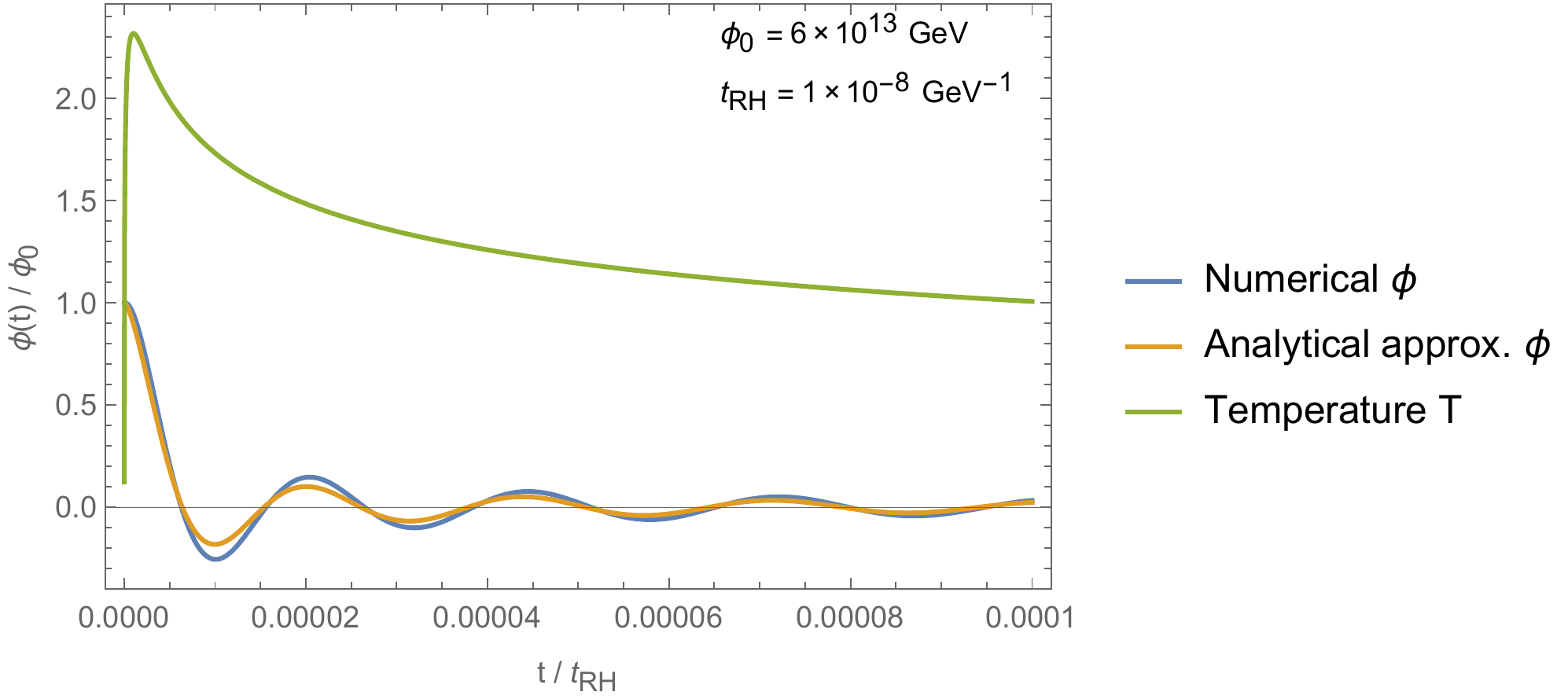}\caption{Higgs evolution with $\Lambda_{I}=1.5\times10^{16} \, \mathrm{GeV}$, $\Gamma_{I}=10^{8} \, \mathrm{GeV}$, and $\phi_{0}=6\times10^{13} \, \mathrm{GeV}$. First crossing time are $6.28\times10^{-14} \, \mathrm{GeV}^{-1}$
(numerical) and $6.39\times10^{-14} \, \mathrm{GeV}^{-1}$ (analytical approximation).}
\label{fig:Higgs-evolution} 
\end{figure}

For $t_{\mathrm{osc}}< t< t_{RH}$, we approximate the temperature of the plasma by
\begin{equation}
T\left(t\right)\simeq T_{RH}\left(\frac{t_{RH}}{t}\right)^{1/4},
\end{equation}
where the reheat temperature is $T_{RH}\approx\left(3/\pi^{3}\right)^{1/4}g_{*S}^{-1/4}(T_{RH})\sqrt{m_{pl}\Gamma_{I}}$ and $t_{RH}=1/\Gamma_{I}$ is the time when reheating is complete.   For times between $t_{\mathrm{osc}}$ and $t_{RH}$, the equation of motion for the scalar field is then
\begin{equation}
\ddot{\phi}\left(t\right) + \frac{2}{t}\dot{\phi}\left(t\right) + \alpha_T^{2}\frac{T_{RH}^{2}\sqrt{t_{RH}}}{\sqrt{t}}\phi\left(t\right) = 0\label{eq:EOM_T2}
\end{equation}
if the thermal corrections dominate the effective potential, and we have taken $H\left(t\right)\approx2/3t$ since during the epoch in which the inflaton undergoes coherent oscillation the universe evolves as if it were matter dominated.  We can rescale $\phi\left(t=xt_{RH}\right)=\phi_{0}y\left(x\right)$ and $t=xt_{RH}$ to rewrite Eq.~\eqref{eq:EOM_T2} as
\begin{equation}
y''\left(x\right)+\frac{2}{x}y'\left(x\right)+\frac{\alpha_T^{2}\beta^{2}}{\sqrt{x}}y\left(x\right)=0,\label{eq:EOM_T2_y(x)}
\end{equation}
where $\beta=T_{RH}t_{RH}=6.06\times10^{4}\left(\frac{10^{8}\text{GeV}}{\Gamma_{I}}\right)^{1/2}$.
The independent solutions for Eq.~\eqref{eq:EOM_T2_y(x)} are
\begin{align}
y_{1}\left(x\right) & =\left(\frac{3}{2}\right)^{2/3}\Gamma\left(\frac{5}{3}\right)J_{2/3}\left(\frac{4\alpha_T\beta}{3}\,x^{3/4}\right)\frac{1}{\left(\alpha_T\beta\right)^{2/3}\sqrt{x}},\label{eq:y1}\\
y_{2}\left(x\right) & =\left(\frac{3}{2}\right)^{2/3}\Gamma\left(\frac{1}{3}\right)J_{-2/3}\left(\frac{4\alpha_T\beta}{3}\,x^{3/4}\right)\frac{1}{\left(\alpha_T\beta\right)^{2/3}\sqrt{x}},
\end{align}
where $J_{n}\left(z\right)$ is the Bessel function of the first kind.
Since $y_{2}\left(0\right)$ diverges, and $y_{1}\left(0\right)=1$
and $y_{1}'\left(0\right)=0$, we should take only $y_{1}$ as the
physical solution, subject to the boundary condition that $\phi(t=0) = \phi_0$ (where we shift our zero of time by $t_\mathrm{osc}$).  Both the analytical solution given by Eq.~\eqref{eq:y1} with this boundary condition and the actual numerical solution are shown in Fig.~\ref{fig:Higgs-evolution}.

As discussed in \cite{Yang:2015ida}, one must be concerned with washout due to the subsequent oscillations of the Higgs VEV.  This is avoided when the scattering processes are not too efficient in the early universe (which gives the result that a large chemical potential is needed to generate the asymmetry).  Washout can be avoided either by having these interactions turn off rapidly, or by considering parameters such that there is significant damping of the Higgs oscillations, such as those in Fig.~\ref{fig:Higgs-evolution}.  Regardless of the balance of factors, the end of the asymmetry production, $t_\mathrm{rlx}$, occurs around the time when the Higgs VEV passes zero.  This can be approximated analytically by noting that the Bessel function with $n=2/3$ has a first zero at $z_{0}=3.376$.  The relaxation time of the scalar field can then be approximated using the first crossing at
\begin{equation}
z_{0}=\frac{4\alpha_T\beta}{3}\,x_{\mathrm{rlx}}^{3/4},
\end{equation}
which gives 
\begin{equation}
t_{\mathrm{rlx}}=t_{RH}x_{\mathrm{rlx}} \approx t_{RH}\left(\frac{3z_{0}}{4\alpha_T T_{RH}t_{RH}}\right)^{4/3}.
\end{equation}
Note that since Eq.~\eqref{eq:EOM_T2} is linear in $\phi$, the relaxation
time is independent of the initial $\phi_{0}$. Hence fluctuations
in $\phi_{0}$ does not affect the relaxation time, in the regime considered here: where the potential of the scalar field is dominated by the thermal mass and $t_\mathrm{rlx} < t_{RH}$.  In fact, as long as the potential is dominated by the thermal mass term (quadratic in $\phi$), the relaxation time is always independent of $\phi_0$.

The final lepton-to-entropy ratio can be estimated by
\begin{equation}
Y \approx \dfrac{45}{2 \pi^2 g_{*S}} \dfrac{2 \phi_0^2}{\pi^2 \Lambda_n^2} \dfrac{T_{\mathrm{rlx}}^2 t_{\mathrm{rlx}} \Gamma_I^2}{T_R^3} \mathrm{min}\left[ 1, \dfrac{2}{\pi^2} \sigma_R T_{\mathrm{rlx}}^3 t_\mathrm{rlx} \right] 
\exp \left( - \dfrac{8 + \sqrt{15}}{\pi^2} 
\dfrac{\sigma_R T_{RH}^3}{\Gamma_I}
\right) ,
\end{equation}
which can be found in \cite{Gertov:2016uzs} and improves on the estimates in \cite{Kusenko:2014lra,Yang:2015ida} by $\mathcal{O}(1)$ factors.  In this expression, $\sigma_R$ is the thermally averaged cross section for the lepton-number-violating interaction, $h^0 \bar{\nu} \leftrightarrow h^0 \nu$, and a thermal distribution has been assumed for participating particles.  Using the above expressions, we have
\begin{equation}
Y \approx \dfrac{90 \sigma_R}{\pi^6 g_{*S}}
\left( \dfrac{\phi_0}{\Lambda_n} \right)^2 
\dfrac{3 z_0 T_{RH}}{4 \alpha_T t_{RH}} 
\exp \left( - \dfrac{8 + \sqrt{15}}{\pi^2} \sigma_R T_{RH}^3 t_{RH} \right).
\label{eq:Y_approx_final}
\end{equation}
Since $T_{RH}$ and $t_{RH}$ are independent of $\phi_0$, Eq.~\eqref{eq:Y_approx_final} entails that $Y \propto \phi_0^2$.  Note that since $t_\mathrm{rlx}$ and therefore $T_\mathrm{rlx}$ are independent of $\phi_0$, this is true whether the scale $\Lambda_n$ in the $\mathcal{O}(6)$ operator \eqref{eq:O6operator} is a constant or whether it is the temperature of the plasma.

\section{Power Spectrum of the Lepton Asymmetry}
\label{ap:power_spectrum_Y}

In the case that a scalar field $\phi\left(x\right)$ has a non-zero homogeneous part,
$\left\langle \phi\left(x\right)\right\rangle \neq0,$ the
fluctuation in any quantity that scales as $X \propto \phi^2 $ is simply $\delta X\propto2\left|\left\langle \phi\left(x\right)\right\rangle\right| \delta \phi$
for small $\delta \phi$, which gives 
\begin{equation}
\dfrac{\delta X}{\left\langle X\right\rangle}\approx2\dfrac{\delta \phi}{\left|\left\langle \phi\left(x\right)\right\rangle \right|}.
\label{dY_linear}
\end{equation}

However, this is not applicable to the baryonic asymmetry in the relaxation leptogenesis model because the homogeneous part
of $\phi$ is zero, $\left\langle \phi\left(x\right)\right\rangle =0$, due to the symmetry of the potential.  We note that it is $\phi_0 \equiv \sqrt{\left\langle \phi^{2}\left(x\right)\right\rangle }$ which is nonzero, and as we have explained in the Appendix \ref{ap:analytical}, the lepton asymmetry depends on the initial value of $\phi$ via $Y\propto\phi_{0}^{2}$.  In this appendix, we now proceed to calculate the primordial power spectrum of the lepton asymmetry taking into account the fact that it is $\sqrt{\left\langle \phi^{2}\left(x\right)\right\rangle }$, not $\left\langle \phi(x) \right\rangle$, which is nonzero.

In the following analysis, we adopt the following conventions for the Fourier transform:
\begin{align}
\phi\left(x\right) & =\int\frac{d^{3}k}{\left(2\pi\right)^{3}}e^{i\vec{k}\cdot\vec{x}}\phi_{\vec{k}},\\
\phi_{\vec{k}} & =\int d^{3}xe^{-i\vec{k}\cdot\vec{x}}\phi\left(x\right).
\end{align}
The power spectrum of $\phi$, $\mathcal{P}_{\phi}\left(k\right)$,
is defined through the two-point correlation function of $\phi_{\vec{k}}$
\begin{equation}
\left\langle \phi_{\vec{k}}\phi_{\vec{k}'}\right\rangle =\left(2\pi\right)^{3}\delta^{3}\left(\vec{k}+\vec{k}'\right)\frac{2\pi^{2}}{k^{3}}\mathcal{P}_{\phi}\left(k\right).\label{eq:2_point_correlation_phi}
\end{equation}
As we mentioned in Eq.~\eqref{eq:Pc_phi(k)}, we approximate the power
spectrum of $\phi$ by
\begin{align}
\mathcal{P}_{\phi}\left(k\right) & =\left(\frac{H_{I}}{2\pi}\right)^{2}\theta\left(k-k_{s}\right)\theta\left(k_{s}e^{N_{\mathrm{last}}}-k\right).\label{eq:power_spectrum_of_phi}
\end{align}
We remind the reader that $k_s$ is the comoving scale which leaves the horizon when the Higgs VEV begins growing, $N_\mathrm{last}$ $e$-folds before the end of inflation.  Our results are insensitive to very large values of $k$; however, for completeness, we have included a high-scale cutoff imposed by the fact that $\phi$ grows until the end of inflation.  The comoving scale $k$ that leaves the scale at the end of inflation is the highest scale on which isocurvature modes are produced; this scale is $k_s e^{N_\mathrm{last}}$.  Again, though, such high $k$ values are not relevant to our results, which means that we are insensitive to the end of inflation.

We now look at the fluctuation of $f\left(x\right)\equiv\phi^{2}\left(x\right)$ with respect to its expectation value,
\begin{equation}
\delta f\left(x\right)=\phi^{2}\left(x\right)-\left\langle \phi^{2}\left(x\right)\right\rangle =\int\frac{d^{3}k}{\left(2\pi\right)^{3}}e^{i\vec{k}\cdot\vec{x}}f_{\vec{k}}.
\end{equation}
The power spectrum of $\delta f\left(x\right)$ can be computed from
the two-point function of the Fourier transform of $\delta f$
\begin{equation}
\left\langle f_{\vec{k}}f_{\vec{k}'}\right\rangle =\left(2\pi\right)^{3}\delta^{3}\left(\vec{k}+\vec{k}'\right)\frac{2\pi^{2}}{k^{3}}\mathcal{P}_{\delta f}\left(k\right),\label{eq:2_point_correlation_phi-1}
\end{equation}
 which is 
\begin{align}
\left\langle f_{\vec{k}}f_{\vec{k}'}\right\rangle  & =\int d^{3}xd^{3}ye^{-i\vec{k}\cdot\vec{x}-i\vec{k}'\cdot\vec{y}}\left\langle \delta f\left(x\right)\delta f\left(y\right)\right\rangle \\
 & =\int d^{3}xd^{3}ye^{-i\vec{k}\cdot\vec{x}-i\vec{k}'\cdot\vec{y}}\left[\left\langle \phi^{2}\left(x\right)\phi^{2}\left(y\right)\right\rangle -\left\langle \phi^{2}\left(x\right)\right\rangle \left\langle \phi^{2}\left(y\right)\right\rangle \right]\\
 & =\int d^{3}xd^{3}ye^{-i\vec{k}\cdot\vec{x}-i\vec{k}'\cdot\vec{y}}\int\frac{d^{3}k_{1}d^{3}k_{2}d^{3}k_{3}d^{3}k_{4}}{\left(2\pi\right)^{12}}e^{i\left(\vec{k}_{1}+\vec{k}_{2}\right)\cdot\vec{x}}e^{i\left(\vec{k}_{3}+\vec{k}_{4}\right)\cdot\vec{y}}\nonumber \\
 & \quad\times\left(\left\langle \phi_{\vec{k}_{1}}\phi_{\vec{k}_{2}}\phi_{\vec{k}_{3}}\phi_{\vec{k}_{4}}\right\rangle -\left\langle \phi_{\vec{k}_{1}}\phi_{\vec{k}_{2}}\right\rangle \left\langle \phi_{\vec{k}_{3}}\phi_{\vec{k}_{4}}\right\rangle \right).
\end{align}
Using Wick's theorem, one can express the 4-point function in terms
of 2-point functions as
\begin{equation}
\left\langle \phi_{\vec{k}_{1}}\phi_{\vec{k}_{2}}\phi_{\vec{k}_{3}}\phi_{\vec{k}_{4}}\right\rangle =\left\langle \phi_{\vec{k}_{1}}\phi_{\vec{k}_{2}}\right\rangle \left\langle \phi_{\vec{k}_{3}}\phi_{\vec{k}_{4}}\right\rangle +\left\langle \phi_{\vec{k}_{1}}\phi_{\vec{k}_{3}}\right\rangle \left\langle \phi_{\vec{k}_{2}}\phi_{\vec{k}_{4}}\right\rangle +\left\langle \phi_{\vec{k}_{1}}\phi_{\vec{k}_{4}}\right\rangle \left\langle \phi_{\vec{k}_{2}}\phi_{\vec{k}_{3}}\right\rangle .
\end{equation}
 Integrating over $\vec{x}$ and $\vec{y}$, and making use of Eq.~\eqref{eq:2_point_correlation_phi}, we have
\begin{align}
\left\langle f_{\vec{k}}f_{\vec{k}'}\right\rangle  & =2\int d^{3}k_{1}d^{3}k_{2}d^{3}k_{3}d^{3}k_{4}\delta^{3}\left(\vec{k}-\vec{k}_{1}-\vec{k}_{2}\right)\delta^{3}\left(\vec{k}'-\vec{k}_{3}-\vec{k}_{4}\right)\nonumber \\
 & \quad\times\delta^{3}\left(\vec{k}_{1}+\vec{k}_{3}\right)\delta^{3}\left(\vec{k}_{2}+\vec{k}_{4}\right)\frac{2\pi^{2}}{k_{1}^{3}}\frac{2\pi^{2}}{k_{2}^{3}}\mathcal{P}_{\phi}\left(k_{1}\right)\mathcal{P}_{\phi}\left(k_{2}\right)\\
 & =2\int d^{3}k_{1}d^{3}k_{2}\delta^{3}\left(\vec{k}-\vec{k}_{1}-\vec{k}_{2}\right)\delta^{3}\left(\vec{k}'+\vec{k}_{1}+\vec{k}_{2}\right)\frac{4\pi^{4}}{k_{1}^{3}k_{2}^{3}}\mathcal{P}_{\phi}\left(k_{1}\right)\mathcal{P}_{\phi}\left(k_{2}\right).\\
 & =2\delta^{3}\left(\vec{k}+\vec{k}'\right)\int d^{3}k_{1}\frac{4\pi^{4}}{k_{1}^{3}\left|\vec{k}_{1}-\vec{k}\right|^{3}}\mathcal{P}_{\phi}\left(k_{1}\right)\mathcal{P}_{\phi}\left(\left|\vec{k}_{1}-\vec{k}\right|\right).
\end{align}
Thus, the power spectrum of $\delta f$ is 
\begin{equation}
\mathcal{P}_{\delta f}\left(k\right)=\frac{k^{3}}{2\pi}\int d^{3}k_{1}\frac{1}{k_{1}^{3}\left|\vec{k}_{1}-\vec{k}\right|^{3}}\mathcal{P}_{\phi}\left(k_{1}\right)\mathcal{P}_{\phi}\left(\left|\vec{k}_{1}-\vec{k}\right|\right).
\end{equation}

\begin{figure}
\includegraphics[width=0.4\paperwidth]{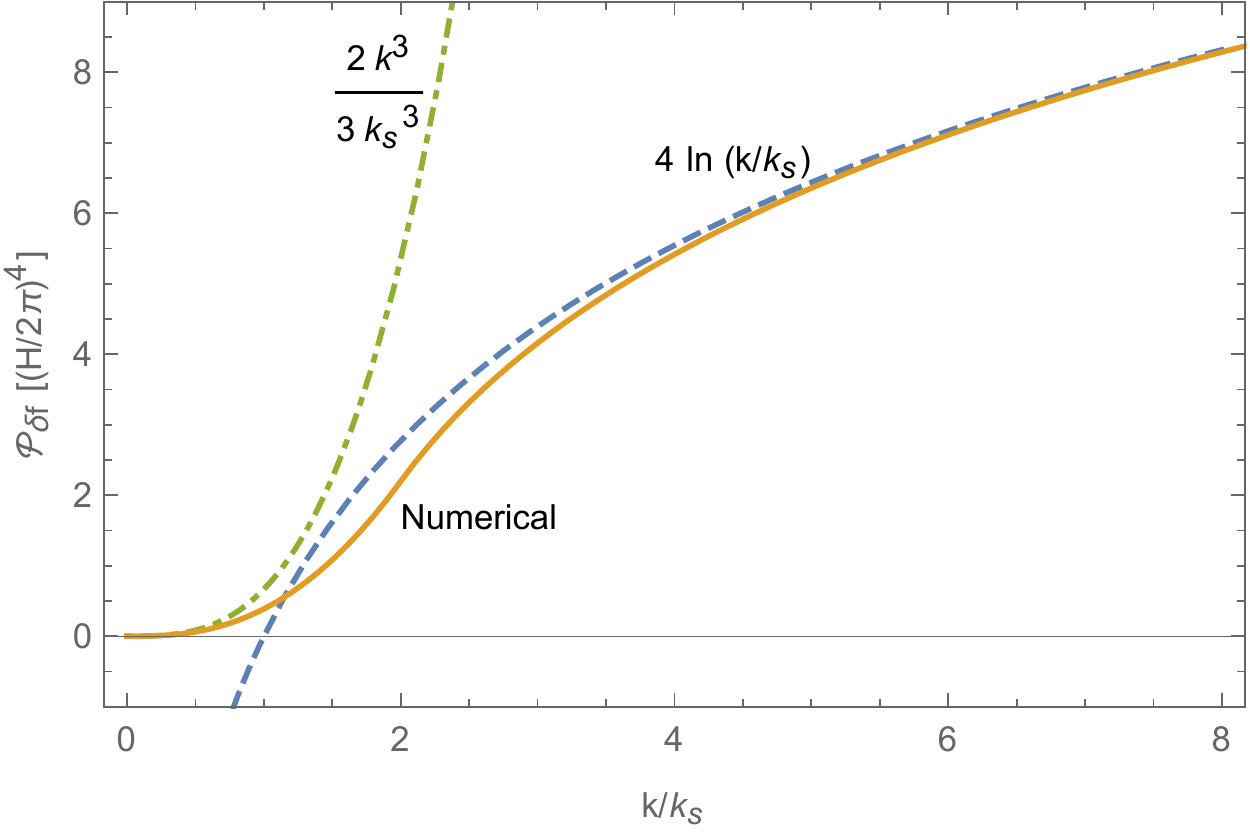}

\caption{Power spectrum of the fluctuation of $f=\phi^{2}$ with respect to
its expectation value, $\left\langle \phi^{2}\right\rangle $. The
yellow solid line denotes the numerical integration result of Eq.~\eqref{eq:Power_spectrum_df_ori}.
The blue dashed curve shows the approximation \eqref{eq:Power_spectrum_df_ks}
for $k>k_{s}$. The green dash-dotted line shows the $\left(k/k_{s}\right)^{3}$
suppression as described by Eq.~\eqref{eq:Power_spectrum_df_small}, for $k<k_{s}$.
\label{fig:Power_spectrum_df_small}}
\end{figure}

\begin{figure}
\includegraphics[width=0.4\paperwidth]{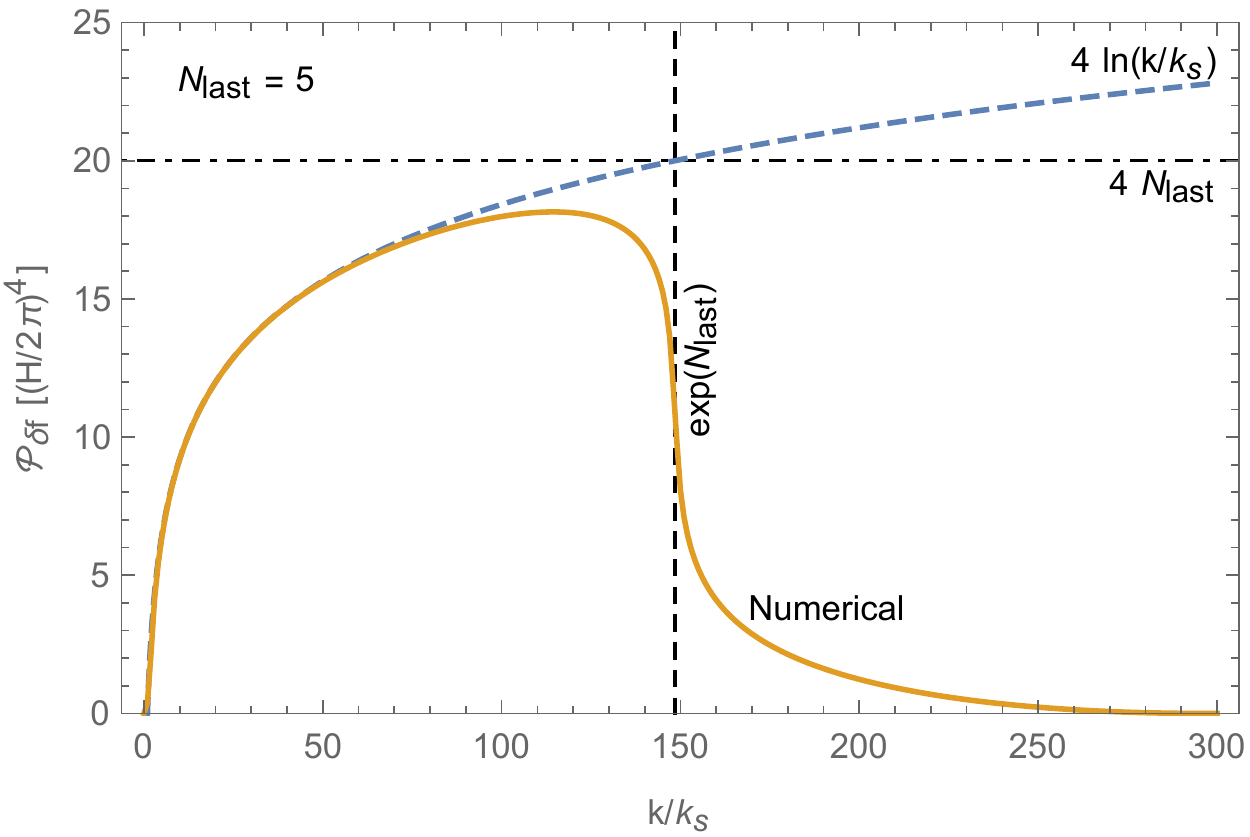}

\caption{Same plot as Fig.~\ref{fig:Power_spectrum_df_small} with $N_{\mathrm{last}}=5$
as an example; this enables us to see the large scale cutoff.  The yellow solid line denotes the numerical result.  The blue dashed curve shows the approximation \eqref{eq:Power_spectrum_df_ks}.
The deviation between them appears at the scale $k\sim k_{s}e^{N_{\mathrm{last}}}$.  The power spectrum reaches an upper limit around $4N_{\mathrm{last}}\left(H_{I}/2\pi\right)^{4}$.  Since our calculation of the CIB is dominated by $k \approx 1.4 k_s$, the large scale cutoff is irrelevant to our signal.
\label{fig:Power_spectrum_df_large}}
\end{figure}

For the power spectrum of $\phi$ given by Eq.~\eqref{eq:power_spectrum_of_phi},
this gives
\begin{align}
\mathcal{P}_{\delta f}\left(k\right) & =\frac{k^{3}}{2\pi}\left(\frac{H_{I}}{2\pi}\right)^{4}\int d^{3}k_{1}\frac{1}{k_{1}^{3}\left|\vec{k}_{1}-\vec{k}\right|^{3}}\theta\left(k_{1}-k_{s}\right)\theta\left(k_{s}e^{N_{\mathrm{last}}}-k_{1}\right)\nonumber \\
 & \quad\theta\left(\left|\vec{k}_{1}-\vec{k}\right|-k_{s}\right)\theta\left(k_{s}e^{N_{\mathrm{last}}}-\left|\vec{k}_{1}-\vec{k}\right|\right).\label{eq:Power_spectrum_df_ori}
\end{align}
For $k\ll k_{s}$, the power spectrum is suppressed as
\begin{equation}
\mathcal{P}_{\delta f}\left(k\right)\approx\frac{k^{3}}{2\pi}\left(\frac{H_{I}}{2\pi}\right)^{4}\int_{k_{s}}^{\infty}\frac{4\pi dk_{1}}{k_{1}^{4}}=\frac{2}{3}\left(\frac{H_{I}}{2\pi}\right)^{4}\left(\frac{k}{k_{s}}\right)^{3}.\label{eq:Power_spectrum_df_small}
\end{equation}
For $k_{s}<k<k_{s}e^{N_{\mathrm{last}}}$, integral is dominated
by $\vec{k}_{1}\sim\vec{k}_{s}$ and $\vec{k}_{1}\sim\vec{k}-\vec{k}_{s}$,
so one can approximate (see also Appendix A of \cite{Lyth:1991ub})
\begin{align}
\mathcal{P}_{\delta f}\left(k\right) & \approx\frac{k^{3}}{2\pi}\left(\frac{H_{I}}{2\pi}\right)^{4}2\int_{k_{s}}^{k}\frac{4\pi k_{1}^{2}dk_{1}}{k_{1}^{3}k^{3}}\\
 & =4\left(\frac{H_{I}}{2\pi}\right)^{4}\ln\left(\frac{k}{k_{s}}\right).\label{eq:Power_spectrum_df_ks}
\end{align}
The power spectrum reaches a maximum $\sim 4N_{\mathrm{last}}\left(H_{I}/2\pi\right)^{4}$ before being suppressed severely beyond $k = k_s e^{N_{\mathrm{last}}}$.  However, as mentioned, this large scale cutoff does not affect our CIB signal, which is dominated by $k \approx 1.4 k_s$.  The behavior of Eqs.~\eqref{eq:Power_spectrum_df_ori}, \eqref{eq:Power_spectrum_df_small},
and \eqref{eq:Power_spectrum_df_ks} are shown in Figs.~\ref{fig:Power_spectrum_df_small}
and \ref{fig:Power_spectrum_df_large}. 

 Since the fluctuation of $\delta f$ is suppressed for $k<k_{s}$,
we take 
\begin{equation}
\mathcal{P}_{\delta f}\left(k\right)\approx4\left(\frac{H_{I}}{2\pi}\right)^{4}\ln\left(\frac{k}{k_{s}}\right)\theta\left(k-k_{s}\right)
\end{equation}
for $k \ll k_s e^{N_{\mathrm{last}}}$.
The average fluctuation of $f$ per $\ln k$
interval is then given by $\delta f_{k}=\sqrt{\mathcal{P}_{\delta f}\left(k\right)}$.
Therefore, the spectrum of the fluctuation of $Y_{B}$ is
\begin{equation}
\left.\frac{\delta Y_{B}}{Y_{B}}\right|_{k}=\frac{\delta f_{k}}{\left\langle f\right\rangle }\approx\frac{2\ln^{1/2}\left(k/k_{s}\right)}{N_{\mathrm{last}}}\theta\left(k-k_{s}\right),
\end{equation}
as used in Eq.~\eqref{eq:YB_ini}.

\bibliographystyle{apsrev4-1}
\bibliography{Reference}

\end{document}